\documentclass[10pt,journal,compsoc]{IEEEtran}

\usepackage{cite}
\usepackage{float}
\usepackage[dvipsnames]{xcolor}
\usepackage[utf8]{inputenc} % allow utf-8 input
\usepackage[T1]{fontenc}    % use 8-bit T1 fonts
\usepackage{hyperref}       % hyperlinks
\usepackage{url}            % simple URL typesetting
\usepackage{booktabs}       % professional-quality tables
\usepackage{amsfonts}       % blackboard math symbols
\usepackage{nicefrac}       % compact symbols for 1/2, etc.
\usepackage{microtype}      % microtypography
\usepackage{tcolorbox}
\usepackage{adjustbox}
\usepackage{multirow}
\usepackage[frozencache,cachedir=.]{minted}
\usepackage{array}
\usepackage{siunitx}
\setminted{fontsize=\tiny}
\definecolor{bg}{HTML}{282828}
\usepackage{pmboxdraw}
\usepackage{algorithmic}
\usepackage{listings}
\usepackage[linesnumbered,ruled,vlined]{algorithm2e}
\def\BibTeX{{\rm B\kern-.05em{\sc i\kern-.025em b}\kern-.08em
    T\kern-.1667em\lower.7ex\hbox{E}\kern-.125emX}}

\usepackage{enumitem}
\usepackage[keeplastbox]{flushend}

\usepackage{flushend}
\usepackage{balance}

\usepackage[justification=centering]{caption}

\begin{document}

\newcommand{\ie}{\textit{i.e.,}~}
\newcommand{\eg}{\textit{e.g.,}~}
\newcommand{\etc}{\textit{etc.}~}
\newcommand{\etal}{\textit{et al.}~}

%Comments
\newcommand{\nb}[2]{
    \fbox{\bfseries\sffamily\scriptsize#1}
    {\sf\small$\blacktriangleright$\textit{#2}$\blacktriangleleft$}
}

\newcommand\MICHELE[1]{\textcolor{blue}{\nb{MICHELE}{#1}}}
\newcommand\DAWN[1]{\textcolor{green}{\nb{DAWN}{#1}}}
\newcommand\ALEXEY[1]{\textcolor{red}{\nb{ALEXEY}{#1}}}

\newcommand{\approach}{{\sc AthenaTest}\xspace}
\newcommand{\dataset}{{\sc Methods2Test}\xspace}

%%% Coloring the comment as blue
\newcommand\mycommfont[1]{\footnotesize\ttfamily\textcolor{blue}{#1}}
\SetCommentSty{mycommfont}

\SetKwInput{KwInput}{Input}                % Set the Input
\SetKwInput{KwOutput}{Output}              % set the Output

% displaying code
\lstdefinestyle{myJavaStyle}{
  frame=tb,
  float=*,
  language=java,
  aboveskip=3mm,
  belowskip=3mm,
  showstringspaces=false,
  columns=flexible,
  basicstyle={\small\ttfamily},
  numbers=none,
  numberstyle=\tiny\color{gray},
  keywordstyle=\color{blue},
  commentstyle=\color{dkgreen},
  stringstyle=\color{mauve},
  frame=single,
  breaklines=true,
  breakatwhitespace=true,
  tabsize=3,
}

\title{Unit Test Case Generation with Transformers\\ and Focal Context}

\author{Michele Tufano, Dawn Drain, Alexey Svyatkovskiy, Shao Kun Deng, Neel Sundaresan% <-this % stops a space
\IEEEcompsocitemizethanks{\IEEEcompsocthanksitem M. Tufano, D. Drain, A. Svyatkovskiy, S. K. Deng, N. Sundaresan are with Microsoft, Redmond, WA, USA.\protect\\
E-mail: \{mitufano, dadrain, alsvyatk, shade, neels\}@microsoft.com}}

% \author{\IEEEauthorblockN{Michele Tufano, Dawn Drain, Alexey Svyatkovskiy, Shao Kun Deng, Neel Sundaresan}
% \IEEEauthorblockA{Microsoft\\
% Redmond, WA, USA\\
% Email: \{mitufano, dadrain, alsvyatk, shade, neels\}@microsoft.com}\vspace{-5ex}}
% \and
% \IEEEauthorblockN{Homer Simpson}
% \IEEEauthorblockA{Twentieth Century Fox\\
% Email: homer@thesimpsons.com}

% \and
% \IEEEauthorblockN{James Kirk\\ and Montgomery Scott}
% \IEEEauthorblockA{Starfleet Academy\\
% San Francisco, California 96678-2391\\
% Telephone: (800) 555--1212\\
% Fax: (888) 555--1212}
% \and
% \IEEEauthorblockN{Homer Simpson}
% \IEEEauthorblockA{Twentieth Century Fox\\
% Email: homer@thesimpsons.com}
% \and
% \IEEEauthorblockN{James Kirk\\ and Montgomery Scott}
% \IEEEauthorblockA{Starfleet Academy\\
% San Francisco, California 96678-2391\\
% Telephone: (800) 555--1212\\

\IEEEtitleabstractindextext{%
\begin{abstract}
Software testing is a critical part of software development life cycle which helps identify potential regressions and reduce maintenance costs, yet it is often neglected by developers. Automated unit test case generation tools facilitate test-driven development and support developers by suggesting tests intended to identify flaws in their code. Existing approaches are usually guided by the test coverage criteria, generating synthetic test cases that are often difficult for developers to read or understand.
In this paper we propose \approach, an approach that aims to generate unit test cases by learning from real-world focal methods and developer-written test cases. We formulate unit test case generation as a sequence-to-sequence learning task, adopting a two-step training procedure consisting of denoising pretraining on a large unsupervised Java corpus, and supervised finetuning for a downstream translation task of generating unit tests. We investigate the impact of natural language and source code pretraining, as well as the focal context information surrounding the focal method. We found that both techniques provide improvements in terms of validation loss, with pretraining yielding 25\% relative improvement and focal context providing additional 11.1\% improvement. 
We also introduce \dataset, the largest publicly available supervised parallel corpus of unit test case methods and corresponding focal methods in Java, which comprises 780K test cases mined from 91K open-source repositories hosted on GitHub.
We evaluate \approach on five defects4j projects, generating $\sim$25K passing test cases covering 43.7\% of the focal methods with only 30 attempts. We execute the test cases, collect test coverage information, and compare them with test cases generated by EvoSuite and GPT-3, finding that our approach outperforms GPT-3 and has comparable coverage w.r.t. EvoSuite. Finally, we survey professional developers on their preference in terms of readability, understandability, and testing effectiveness of the generated test cases. The results show that developers overwhelmingly prefer test cases generated by \approach over EvoSuite's, suggesting that our approach could significantly improve developer productivity.

\end{abstract}

}

\maketitle

%   <concept>
%       <concept_id>10011007.10011074.10011099.10011102.10011103</concept_id>
%       <concept_desc>Software and its engineering~Software testing and debugging</concept_desc>
%       <concept_significance>500</concept_significance>
%       </concept>
%   <concept>
%       <concept_id>10010147.10010178.10010179.10010180</concept_id>
%       <concept_desc>Computing methodologies~Machine translation</concept_desc>
%       <concept_significance>300</concept_significance>
%       </concept>
%  </ccs2012>

% \ccsdesc[500]{Software and its engineering~Software testing and debugging}
% \ccsdesc[300]{Computing methodologies~Machine translation}

% \keywords{software testing, unit test, neural networks}

\section{Introduction}
Software testing is widely acknowledged as one of the most critical, challenging, and expensive phases of the software development lifecycle. Technology companies are constantly looking into ways to deliver their software faster, without sacrificing its quality and correctness. To succeed, these companies often rely on continuous integration and delivery of software, which allows for fast and reliable deployment of software into production. In this context, automated testing represents a fundamental piece of the pipeline, providing developers with the confidence they need to iterate quickly, and integrate new features without regressions.

%In what has become one of the most famous testing concept, Mike Cohn represented \textit{unit testing} as the foundational basis of the software \textit{test pyramid}, beneath integration and end-to-end testing \cite{cohn2010succeeding}.

\textit{Unit testing} lays as the foundational basis of the testing pyramid, beneath integration and end-to-end testing \cite{cohn2010succeeding}. This prominent visual metaphor intends to provide a guidance on the adequate amount of effort that should be allocated for each of the testing layers. Thus, the largest amount of tests should be at the unit test layer, where individual units of software (\eg a single method) are tested in isolation to ensure that they behave as intended. 

Unit Test frameworks, such as JUnit~\cite{junit}, offer an environment and APIs that facilitate writing and executing repeatable test cases. JUnit provides methods such as \textit{assertions} which support the developers in checking conditions, outputs, or states in a software program, assessing its expected behavior. Several other frameworks have been built on top of JUnit, such as Cactus\cite{cactus} and TestnNG\cite{testng}. Others can be integrated with JUnit to support different scenarios or testing methodologies, such as Mockito~\cite{mockito}, which allows mocking of objects by replacing functionalities with dummy implementations that emulate real code, focusing the testing on the method under test.

On top of these frameworks, researchers have proposed several techniques that aim to automate the generation of unit test cases. EvoSuite~\cite{fraser2011evosuite}, Randoop~\cite{pacheco2007randoop}, and Agitar~\cite{agitar} are among the most popular and widely used examples of such techniques. EvoSuite relies on an evolutionary approach based on a genetic algorithm to generate unit test cases, targeting code coverage criteria such as branch and line coverage. Specifically, it introduces mutants (\ie modified versions of methods or classes under test) and iteratively generates assert statements to kill such mutants. During this process, EvoSuite minimizes the number of asserts while trying to maximize the number of detected mutants. Randoop is a different automated test generation tool that relies on feedback-directed random testing, a technique that uses execution traces to guide the selection of method sequences which are then checked against a set of user-specified contracts (\ie user-specified program logic).

\begin{figure*}[ht]
    \centering
    %\vspace{-0.2cm}
    \includegraphics[width=1.02\textwidth]{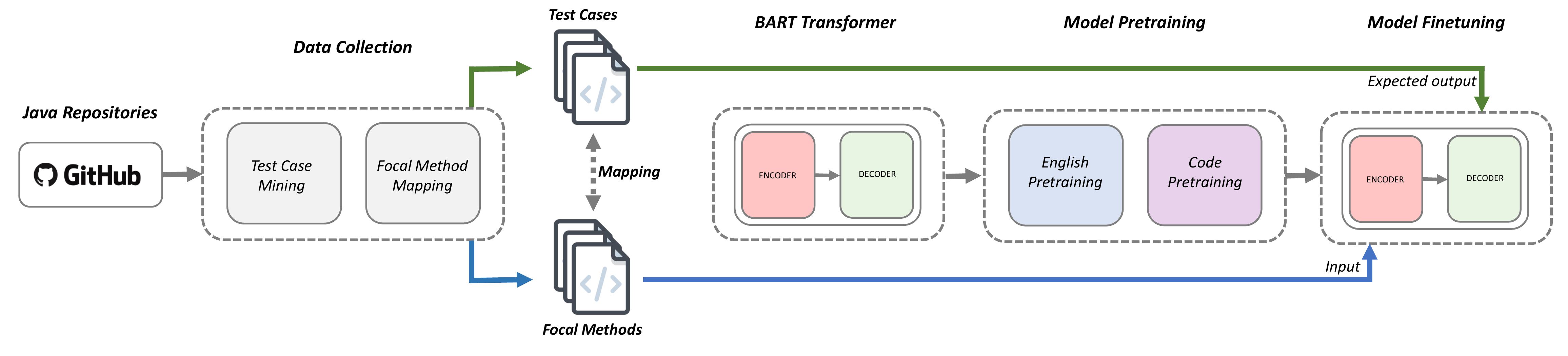}
    \vspace{-0.6cm}
    \caption{Overview of \approach~ -- We mine test cases from GitHub and map them to the corresponding focal methods, which we collect in \dataset, then pretrain a BART Transformer model on both English and Source Code corpora, finally we finetune the model on the unit test case generation task.}
    \vspace{-0.3cm}
    \label{fig:overview}
\end{figure*}

A major weakness and criticism of these approaches is related to the poor readability and understandability of the generated test cases \cite{daka2015modeling, grano2018empirical}, which clearly appear as machine-generated code. Other studies have highlighted different limitations of these automation tools, such as unsatisfactory code quality \cite{palomba2016diffusion, palomba2016automatic, grano2019scented}, poor fault-detection capability \cite{pinto2010multi}, and the inability to adequately meet the software testing needs of industrial developers \cite{almasi2017industrial, shamshiri2015automated}. These limitations stem from the fact that these approaches mainly focus on code coverage as unique objective, disregarding other factors that may be relevant for developers.

Deep learning techniques have shown the potential of learning from real-world examples, and have been employed in several software engineering tasks, such as code completion \cite{svyatkovskiy2019pythia}, automated patch generation \cite{tufano2019empirical, chen2019sequencer}, comment generation \cite{hu2018deep}, and many others \cite{watson2020learning}. Recent advancements in transformer models, such as OpenAI's GPT-3~\cite{brown2020language}, have made headlines and shown impressive results in realistic text generation and question answering tasks.

In this paper, we present an approach that aims to \textit{\textbf{learn from developer-written test cases how to generate correct and readable tests.}} Our approach relies on a large sequence-to-sequence transformer model pretrained both on English and Java source code, then finetuned on the task of generating unit test cases. For this task, we mine thousands of real-world test cases and map them to the corresponding focal methods, then use this parallel corpus for training and evaluation.

To summarize, our contributions are as follows:
\begin{itemize}
    \item \approach: an automated test case generation approach based on a sequence-to-sequence transformer model. The approach is able to generate thousands of syntactically correct, compilable, and passing test cases for Defects4j projects, that invoke a variety of testing APIs. The generated test cases have comparable test coverage w.r.t. EvoSuite and they are preferred by professional developers in terms of readability, understandability, and testing effectiveness. These test cases appear to be: (i) \textit{realistic} -- similar to developer-written test cases; (ii) \textit{accurate} -- correctly asserting the expected behavior of a focal method; (iii) \textit{human-readable} --  readable and understandable code, with good variable and method names.
    
    \item \dataset: the largest publicly available\footnote{\url{https://github.com/microsoft/methods2test}\vspace{-0.8cm}} parallel corpus of test cases mapped to the corresponding focal methods\cite{methods2test}. This dataset enlists 780K mapped test cases, extracted from 91K open source Java projects. 
\end{itemize}

%Will add for the preprint
%

\section{Approach}
Figure \ref{fig:overview} provides an overview of our approach. Starting with a dataset of Java open-source projects obtained from GitHub, we mine test cases and map them to the corresponding focal methods (Sec. \ref{sec:data_collection}). Next, we consider a transformer model (Sec. \ref{sec:transformer}), which has been pretrained on English and source code corpora (Sec. \ref{sec:pretraining}), select the best focal context surrounding the focal method (Sec. \ref{sec:focal_context}), and finetune for the task of generating unit test cases (Sec. \ref{sec:finetuning}).

\subsection{Data Collection}
\label{sec:data_collection}
The goal of this stage is to mine test cases and their corresponding focal methods (\ie the method tested by the test case) from a set of Java projects. We select a 91K sample of all the public GitHub Java repositories declaring an open source license, which have been updated within the last five years, and are not forks.

%Algorithm \ref{alg:data_collection} shows the steps we perform for each project. 
First, we parse each project to obtain classes and methods with their associated metadata. Next, we identify each test class and its corresponding focal class. Finally, for each test case within a test class, we map it to the related focal method obtaining a set of mapped test cases.

\subsubsection*{Parsing}
We parse each project under analysis with the \texttt{tree-sitter} parser\cite{treesitter}. During the parsing, we automatically collect metadata associated with the classes and methods identified within the project. Specifically, we extract information such as method and class names, signatures, bodies, annotations, and variables. The parsed code will be used to identify test cases and corresponding focal methods, as well as augmenting the focal methods with focal context.

\subsubsection*{Find Test Classes}
In this stage, we identify all the test classes, which are classes that contain a test case. To do so, we mark a class as a test class if it contains at least one method with the \texttt{@Test} annotation. This annotation informs JUnit that the method to which it is attached can be run as a test case.

\subsubsection*{Find Focal Classes}
For each test class we aim to identify the focal class which represents the class under test. To this aim, we employ the following two heuristics, in sequence:

\begin{itemize}
\item \textit{Path Matching}: best practices for JUnit testing suggests placing code and corresponding test cases in mirrored folder structure. Specifically, given the class \texttt{src/main/java/Foo.java} the corresponding JUnit test cases should be placed in the class \texttt{src/test/java/FooTest.java}. Our first heuristic tries to identify the folder where the focal class is defined, by following the path of the test class but starting with the \texttt{src/main} folder (\ie production code).

\item \textit{Name Matching}: the name of a test class is usually composed of the name of the focal class, along with a "Test" prefix or suffix. For example, the test case for the class \texttt{Foo.java} would probably be named \texttt{FooTest.java}. Thus, following the path matching heuristic, we perform name matching to identify the focal class by matching the name of the test case without the (optional) "Test" prefix/suffix.
\end{itemize}

%Removing this algo to save space

% \begin{algorithm}[]
% \small
% \DontPrintSemicolon
  
%   \KwInput{Project to analyze: \textit{project}}
%   \KwOutput{Mapped Test Cases: \textit{mappedTestCases}}
  
%   \BlankLine
%   \tcp{Parse project}
%   \textit{classes, methods} $\gets$ parse(\textit{project})

%   \BlankLine
%   \tcp{Identify Test Classes}
%   \textit{testClasses} $\gets$ findTestClasses(\textit{classes})
  
%   \BlankLine
%   \For{$tClass$ \textbf{in} $testClasses$}{
%     \textit{class} $\gets$ findFocalClass($tClass$, $classes$)
%     \BlankLine
    
%     \tcp{Map each Test Case to a Focal Method}
%     \For{$testCase$ \textbf{in} $tClass.methods$}{
%         \textit{method} $\gets$ findFocalMethod($testCase$, $class.methods$)
%         \BlankLine
%         $pair \gets$ new MappedTestCase($testCase$, $method$)
        
%         $mappedTestCases$.add($pair$)
%     }
%   }

%   \Return $mappedTestCases$

% \caption{Extracting Mapped Test Cases}
% \label{alg:data_collection}
% \end{algorithm}

\subsubsection*{Find Focal Method}
For each test case (\ie method within a test class with the \texttt{@Test} annotation) we attempt to identify the corresponding focal method within the focal class. To this aim, we employ the following heuristics:

\begin{itemize}
\item \textit{Name Matching}: following the best practices for naming classes, test case names are often similar to the corresponding focal methods. Thus, the first heuristic attempts to match the test cases with a focal method having a name that matches, after removing possible \texttt{Test} prefix/suffix. 

\item \textit{Unique Method Call}: if the previous heuristic did not identify any focal method, we compute the intersection between (i) the list of method invocations within the test case and (ii) the list of methods defined within the focal class. If the intersection yields a unique method, then we select the method as the focal method. The rationale behind this approach is as follows: since we have already matched the test class with the focal class (with very high confidence heuristics), if the test case invokes a single method within that focal class, it is very likely testing that single method.

\end{itemize}

\subsubsection*{Mapped Test Cases}
The result of the data collection phase is a set of mapped test cases, where each test case is mapped to the corresponding focal method. It is important to note that we discard test cases for which we were not able to identify the focal method using our heuristics. We designed these heuristics to be based on testing best practices, and obtain a correct mapping with very high confidence. This allows us to train our model on test cases that follow best practices, and likely excluding test cases that have been automatically generated.

We collect an initial set of 887,646 mapped test case pairs. From this set, we exclude duplicates, remaining with a total of 780,944 unique mapped test case pairs. Next, we split the dataset into training ($\sim$80\% - 624,022 pairs), validation ($\sim$10\% - 78,534 pairs), and test ($\sim$10\% - 78,388 pairs) sets. We performed this split by carefully taking into account possible data leakage. Specifically, during the split we enforce the constraint that any two data points belonging to the same repository cannot be placed in two different sets (\eg one in training and the other in test). That is, all the data points belonging to the same repository will be placed in the same set. 

Table \ref{tab:methods2test} reports the details of the dataset split, with number of repositories and mapped test cases. The set of mapped test cases will be used to train our model to generate a test case given the focal method. We publicly release the dataset \dataset \cite{methods2test}.

\begin{table}[t]
\centering
	\vspace{0.2cm}
	\caption{\dataset Dataset}
	\label{tab:methods2test}
	\resizebox{0.8\linewidth}{!}{
\begin{tabular}{lrr}
\toprule
Set & Repositories & Mapped Test Cases\\
\midrule
Training & 72,188 & 624,022 \\
Validation & 9,104 & 78,534 \\
Test & 10,093 & 78,388 \\
\midrule
Total & 91,385 & 780,944 \\
\bottomrule
\end{tabular}}
\vspace{-0.4cm}
\end{table}

\smallskip

% -------------------------------------------------------------------------------------

\subsection{BART Transformer}
\label{sec:transformer}
\approach is based on a BART transformer model.
%Overview of BART Transformer
BART~\cite{lewis2019bart} is a denoising autoencoder which utilizes the standard sequence-to-sequence transformer architecture from~\cite{DBLP:journals/corr/VaswaniSPUJGKP17}, substituting ReLUs with GeLU activation functions.

% Why we chose BART
We select the BART model architecture because it facilitates finetuning for the downstream translation task of test case generation, providing a more advanced set of noising transformations, which include token masking, token deletion, infilling and statement permutation. The model is pretrained by corrupting documents and optimizing the cross-entropy loss between the decoder’s output and the original input sequence. 

%details on the model size and configs
We pretrain the BART large model architecture, which has 12 encoder layers and 12 decoder layers. The model is trained in mixed-precision, using Adam stochastic optimization procedure with $\epsilon=10^{-6}$, and $\beta_1=0.9$, $\beta_2=0.98$ optimizer parameters; we apply inverse square root learning rate schedule with the base learning rate of 0.0001, a warmup period of 5000 update steps, and local gradient accumulation with a frequency of 4 update steps.

% -------------------------------------------------------------------------------------

\subsection{Pretraining}
\label{sec:pretraining}
We employ two pretraining stages: English Pretraining, where we perform semi-supervised pretraining on a large corpus of English text, and Code Pretraining, where the model is pretrained on Java source code.

\subsubsection*{English Pretraining}
\label{sec:english_pretraining}
In this stage we pretrain a model in a semi-supervised fashion on a large corpus of English text, with the goal of learning semantic and statistical properties of natural language. The pretraining is performed for 40 epochs on 160GB of English text extracted from books, Wikipedia, and news articles~\cite{liu2019roberta}.

BART is trained in an unsupervised manner. Given corrupted text, its objective is to reconstruct the original text. The particular type of noise used in this work involves masking 30\% of all tokens, with masks covering spans of tokens with lengths following a Poisson distribution parameterized by $\lambda=3$, as well as permuting all sentences.

\subsubsection*{Code Pretraining}
\label{sec:code_pretraining}
In this stage we pretrain a model on source code corpus written in Java language, with the goal of learning syntax and properties of source code.

We collect this code corpus dataset by crawling all public, non-fork Java repositories on GitHub with at least 50 stars. We then deduplicate at the file-level using a hash function. After filtering for permissive licenses and filtering out based on heuristics like the fraction of non-ASCII characters, we are left with 25GB of training data from the 26,000 repositories. For pretraining validation, we use the 239 test Java repositories from the CodeSearchNet \cite{husain2019codesearchnet}, which comprise 600MB. 

A similar pretraining strategy to English pretraining is employed. The source code files are corrupted by deleting 20\% of all tokens independently and rotating half of all documents. This pretraining is performed for 10 epochs.

\subsubsection*{Model Pretraining Variants}
\label{sec:models}
At the end of these stages, we obtain four different variants of the model, based on the level of pretraining performed:
\begin{itemize}
    \item \textit{BART\_Scratch}: a model which has not been pretrained on any corpus but directly finetuned on the test case generation task.
    \item \textit{BART\_English}: a model which has been pretrained on the English corpus and then finetuned for the test case generation task.
    \item \textit{BART\_Code}: a model pretrained on the source code corpus, then finetuned on the test case generation task.
    \item \textit{BART\_English+Code}: a model pretrained first on English and further pretrained on source code corpus, then finetuned on the test case generation task.
\end{itemize}

% -------------------------------------------------------------------------------------

\subsection{Focal Context}
\label{sec:focal_context}
In this section we describe the code representation we build for the input to the model. The goal of this phase is to construct an input which contains the necessary information that the model can leverage to generate correct and useful test cases. Intuitively, the focal method (\ie the method under test) represents the core information to feed to the model. However, additional contextual information can provide important clues for the model to better understand the focal method nature and its context, improving the likelihood of generating test cases that compile and properly test the focal method.

We build different versions of the code input representation -- with diverse degree of focal context -- with the aim of empirically evaluating these code representations. We begin with the core information (\ie focal method) and iteratively add contextual information such as class name, constructors, other method signatures, and fields.

\begin{figure}[t!]
    \centering
    \caption{Focal Context}
    \vspace{0.1cm}
    \includegraphics[width=0.47\textwidth]{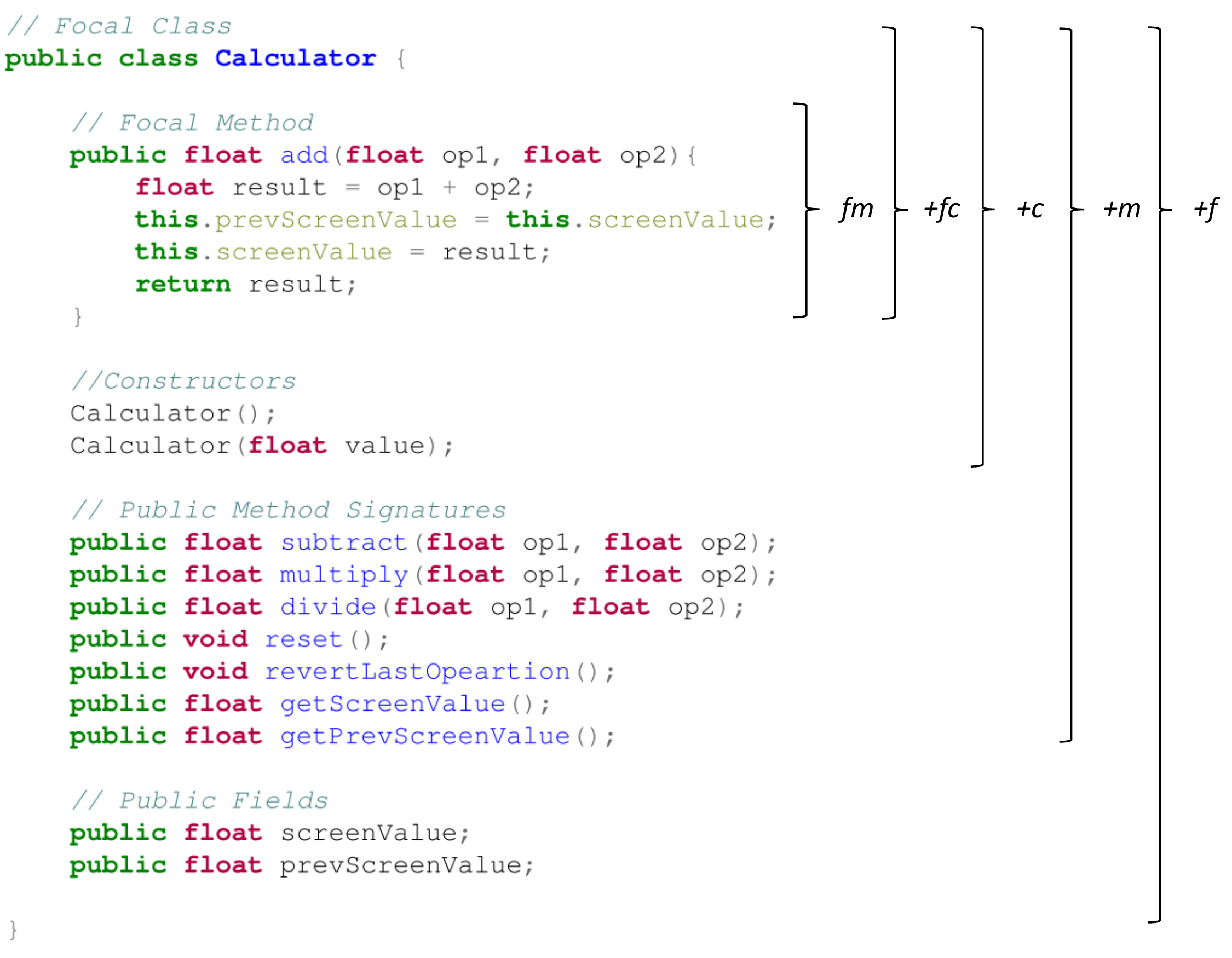}
    \vspace{-0.5cm}
    \label{fig:focal-context}
\end{figure}

Figure \ref{fig:focal-context} provides an overview of the different levels of context we generate for the focal method \texttt{add} in the \texttt{Calculator} class. The left side corresponds to the textual representation, while the right side delineates the context which is indicated with a focal context ID, which we describe in the following:

\begin{itemize}
    \item \textit{fm}: this representation incorporates exclusively the source code of the focal method. Intuitively, this contains the most important information for generating accurate test cases for the given method.
    
    \item \textit{fm+fc}: this representations adds the focal class name, which can provide meaningful semantic information to the model.
    
    \item \textit{fm+fc+c}: this representation adds the signatures of the constructor methods of the focal class. The idea behind this augmentation is that the test case may require instantiating an object of the focal class in order to properly test the focal method.
    
    \item \textit{fm+fc+c+m}: this representation adds the signatures of the other public methods in the focal class. The rationale which motivated this inclusion is that the test case may need to invoke other auxiliary methods within the class (\eg getters, setters) to set up or tear down the testing environment.
    
    \item \textit{fm+fc+c+m+f}: this representation adds the public fields of the focal class. The motivation is that test cases may need to inspect the status of the public fields to properly test a focal method.
\end{itemize}

% We may remove this paragraph below:
 While constructing these representations we face two opposing goals: (i) include as many tokens as possible, given their powerful expressiveness discussed above (ii) keep a concise representation that fits into GPU memory.

Intuitively, having a representation that includes many tokens from the focal context allows the model to \textit{attend} to different parts of the input and leverage these information to generate a correct and meaningful test case. On the other hand, irrelevant tokens could represent noise for the learning process, which could lead to worse performances, as well as wasting GPU memory that could be use for more informative tokens.

It is important to highlight that in our representation, the order of inclusion of a particular focal context, for example the constructors' signatures (\textit{fm+fc+c}) before other methods' signatures (\textit{fm+fc+c+m}), is important, since the textual representation could be truncated if it exceeds 1024 tokens (\ie maximum sequence length in our model). 

This order of inclusion has been defined by the authors based on their understanding and intuition of the meaningful clues for test case generation within the focal class. We empirically evaluate these design decision in our empirical study.

\subsubsection*{Model Context Variants}
Similarly to what has been discussed for model pretraining, we train different model variants for each level of focal context. Specifically, we obtain five different models which we refer to with the corresponding focal context ID.

% ---------- Focal Context Code ----------------
% \begin{figure}[t]
% \vspace{-0.2cm}
%     \centering
% \begin{adjustbox}{width=0.5\textwidth}
% \footnotesize
% \begin{tabular}{c}
% \toprule
%   Focal Method \\
% \midrule
% \begin{minipage}[t]{0.45\textwidth}
% \begin{minted}[escapeinside=||]{java}
% // Focal Class
% public class Calculator {
    
%     // Focal Method
%     public float add(float op1, float op2){
%         float result = op1 + op2;
%         this.prevScreenValue = this.screenValue;
%         this.screenValue = result;
%         return result;
%     }                                       
    
%     //Constructors
%     Calculator();
%     Calculator(float value);
    
%     // Public Method Signatures
%     public float subtract(float op1, float op2);
%     public float multiply(float op1, float op2);
%     public float divide(float op1, float op2);
%     public void reset();
%     public void revertLastOpeartion();
%     public float getScreenValue();
%     public float getPrevScreenValue();
    
%     // Public Fields
%     public float screenValue;
%     public float prevScreenValue;

% }
% \end{minted}
% \end{minipage}\\
% \bottomrule
% \end{tabular}
% \end{adjustbox}
% %\medskip
% \vspace{-0.0cm}
% \caption{Test Cases Generated for \texttt{createFloat}}
% \vspace{-0.1cm}
% \label{fig:new}
% \end{figure}
% ---------- END Code Example ----------------

\subsection{Finetuning}
\label{sec:finetuning}

In this stage we finetune a model on the task of generating unit test cases for a given method. Specifically, we represent this task as a \textit{translation} task, where the source is a focal method (\ie the method we would like to test), and the target is the corresponding test case originally written by a software developer.

The finetuning training is performed using the collected mapped test cases (Sec. \ref{sec:data_collection}), where a mapped test case $mtc_i$ can be seen as a pair $mtc_i = \{tc_i, fm_i\}$ comprising the test case $tc_i$ and the corresponding focal method $fm_i$. The finetuning process is a translation task, with a training objective to learn the mapping $fm_i \rightarrow tc_i$ as a conditional probability $P(tc_i \vert  fm_i)$. Note that we refer with $fm_i$ to the focal method and its available focal context, depending on the model variant.

During training, we use the cross entropy loss and the Adam optimizer, monitoring the loss on the validation set for early stopping. We use shared vocabulary embeddings between Encoder and Decoder for optimization reasons \cite{DBLP:journals/corr/VaswaniSPUJGKP17, DBLP:journals/corr/PressW16} and because our input and output language is the same (\ie Java source code).

%\newpage

\section{Experimental Design}
The goal of our empirical study is to determine if our approach can generate accurate and useful unit test case given a method. Our experiments aim at answering the research questions described in the following paragraphs.

We begin by evaluating the impact of English and code pretraining to select our base model (RQ$_1$), next we consider different levels of focal context and select the best model for unit test generation (RQ$_2$). \approach is then evaluated in terms of code-specific metrics for the test cases it generates (RQ$_3$) and a large-scale analysis is performed on Defects4j projects (RQ$_4$). Finally, we compare \approach against EvoSuite and GPT-3 in terms of code coverage (RQ$_5$) and in terms of developers' preferences (RQ$_6$).

\textbf{RQ$_1$: Does model pretraining impact the performances of Unit Test Case Generation?}
As a first step towards the creation of a unit test generation model, we intend to select a \textit{base} model that we will specialize on our downstream task. The available options (described in Sec. \ref{sec:pretraining}) include a scratch model (randomly initialized, with no pretraining), and model variants with English and/or code pretraining. 

In this research question we aim at evaluating the impact of the pretraining process on the performances of our downstream task. With this aim, we finetune the four model variants on the unit test generation task, letting the models converge independently till no major improvements on the validation loss is observed, and for a maximum of 50k steps. The fintuning at this stage is performed using the minimal level of focal context ($fm$) for all the model variants, since we are only interested in observing the pretraining effect at this point.

We evaluate the models by observing the validation loss during model training. A low validation loss means the model is effectively learning meaningful representations during training and is able to \textit{generalize} how to generate test cases on a different set of input methods (\ie validation set). Specifically, we analyze three key metrics: (i) the initial validation loss during finetuning, which indicates the impact of the pretraining process; (ii) the best validation loss, which highlights the model achieving the best performance; (iii) the number of steps needed to reach the best validation loss, as a measure of how fast the finetuning process converges.

At the end of this experiment, we select the model with the best validation loss, which will be used for further investigation in the following research questions.

\textbf{RQ$_2$: How does focal context impact the training for Unit Test Case Generation?}

In this research question we aim at empirically evaluating the impact of the focal context to the performances of our models on the unit test case generation task. Specifically, the goal is to quantify the effect of each level of focal context, which we add incrementally starting from the focal method. To do so, we perform a preliminary token-based analysis as well as validation loss comparison among the model variants.

\subsubsection*{Ingredient Space Analysis}
Unit test cases may contain tokens that are shared with the focal context, such as variable names, method calls, literals, and so on. We refer to such tokens as \textit{ingredients} that can be selected from the focal context to build a test case candidate. This metaphor has also been used in the literature to characterize tokens necessary to perform bug-fixing activities \cite{martinez2014fix, white2019sorting}.

In order to understand whether different levels of focal context provide possibly more ingredients that the model can leverage to generate a test case, we perform an ingredient space analysis. Specifically, given a focal method \textit{fm}, its corresponding five different levels of focal context (\ie \textit{fm, fm+fc, fm+fc+c, fm+fc+c+m, fm+fc+c+m+f}), and the target test case \textit{tc}, we compute the overlap between the set of tokens in the \textit{tc} and each of the focal context. During this process we properly tokenize the source code and disregard Java keywords and separators. We compare the distributions of number of shared tokens over the training set for the five variants of focal context.

\subsubsection*{Validation Loss}
While a token-based analysis can provide meaningful evidence to support the decision to incorporate a particular focal context, such an analysis is limited in its nature, since it requires perfect token matching. On the other hand, some tokens carry significant semantic value that can provide powerful clues to the generation of test cases, even whew such token does not appear in the test case. For example, the name of the focal class \texttt{Calculator} provides to the model the domain where the focal method belongs to, even in the case that the token \texttt{Calculator} is never used in the actual test case.

For this reason, we complement the ingredient space analysis with a validation loss analysis, where we train five models to generate test cases, each of them taking as input a different version of focal context. Note that in this experiment, while the input source is different, the output target is the same and consistent among the variants. The training is performed starting, for all the variants, from the pretrained model that achieved the best results in the first research question.

\smallskip

At the end of this research question, the results of the analyses will inform us on the variant of focal context that makes the best use of the limited token window (\ie 1024 tokens). This model is the final model for \approach which will be deployed and evaluated in the subsequent research questions.

\textbf{RQ$_3$: What is the quality of the generated Test Cases?}
In this research question we further analyze the test cases generated by the model selected in RQ$_2$. The focus of this analysis is to scrutinize the generated model's predictions looking for specific properties that unit test cases should have.

\subsubsection*{Syntactic Correctness}
We begin by verifying that the sequence of tokens generated by the model represents a syntactically correct source code method conforming to the Java specifications. To this aim, we parse all the predictions generated by the model using a Java parser, which determines the syntactic correctness.

\subsubsection*{Testing APIs}
For a method to be considered as a test case, it needs to exhibit some basic properties, such as: 

\begin{itemize}
    \item \textit{Test Annotation}: the test case should declare the \texttt{@Test} annotation.
    \item \textit{Focal Method Invocation}: to properly test a focal method, the test case should invoke the focal method.
    \item \textit{Testing APIs}: the test case should check the proper behavior of the focal method using testing APIs, such as assert statements and mocking methods. Specifically, we consider two testing framework APIs: JUnit Assert APIs (\eg \texttt{assertTrue}, \texttt{assertEqual}) as well as the Mockito Framework APIs (\eg \texttt{mock}, \texttt{verify}). We chose these testing framework for their popularity and applicability in many different contexts and domains. We plan to incorporate more domain-specific testing frameworks, such as Selenium~\cite{bruns2009web} or REST Assured~\cite{rest-assured} in future work.
\end{itemize}

We check compliance to these properties using a Java parser, extracting annotations and method calls. We also compare the distribution of testing APIs between the original test cases and the ones generated by the model.

\smallskip

\textbf{RQ$_4$: Can \approach generate Test Cases for Defects4j projects?}

In this research question we are interested in evaluating the performances of \approach on a widely common benchmark dataset such as defects4j. We rely on defects4j since it provides a reliable infrastructure to generate, compile, execute, and evaluate test cases for several popular open source software projects.

The goal of this research question is to understand the real-world performance of our approach when used on large and complex systems. In particular, whether \approach is able to generate test cases that are compilable, executable, and correct w.r.t. the given project and focal method. 

Table \ref{tab:defects4j} lists the defects4j projects representing the scope of our empirical analysis. Specifically, we select five popular and commonly used projects: Apache Commons Lang \cite{apachelang}, JFreeChart \cite{JFreeChart}, Apache Common Cli \cite{apachecli}, Apache Common Csv \cite{apachecsv}, Google Gson \cite{googlegson}. We selected these projects as representative of different domains, sizes, and organizations.

Our experimental design consists of three main phases: (i) generation; (ii) execution; and (iii) evaluation of the test cases.

\subsubsection*{Generation}
For each project $p$ and revision $rev$, we checkout the fixed version $rev_f$, since in our experimental scenario we are generating test cases assuming a correct project. Next, we identify the focal class(es) $fc$ (\ie the class where the bug was identified) for the specific $rev$ using defects4j APIs. Subsequently, we parse the focal classes and extract the list of every public method. Each one of these public methods represents a focal method $fm$ for which we aim to generate test cases. For each focal method $fm$ we invoke \approach and generate 30 candidate test cases using beam search.

\subsubsection*{Execution}
Each candidate test case $tc$ is then injected into a test class that contains the appropriate imports and scaffolding necessary to be executed for the particular project $p$ and revision $rev$. Next, the test class is compressed into a format supported by defects4j API and the test case is executed. During execution we collect coverage information with Cobertura, which produces an xml file specifying the lines and conditions covered for each Java file and method.

\begin{table}[t]
\centering
	\vspace{0.2cm}
	\caption{Defects4j Projects Analyzed}
	\label{tab:defects4j}
	\resizebox{0.7\linewidth}{!}{
\begin{tabular}{lrr}
\toprule
Project & Revisions & Focal Methods\\
\midrule
Lang & 63 & 2,712 \\
Chart & 26 & 1,328 \\
Cli & 38 & 645 \\
Csv & 16 & 373 \\
Gson & 18 & 220 \\
\midrule
Total & 161 & 5,278 \\
\bottomrule
\end{tabular}}
\vspace{-0.5cm}
\end{table}

\subsubsection*{Evaluation}
After the execution phase, we perform the evaluation by analyzing multiple output files and logs, including build logs, test execution outputs, and coverage files.

We classify each candidate test case $tc$ into these categories:

\begin{itemize}
    \item \textit{Syntax Error}: the test has syntax errors;
    \item \textit{Build Error}: the test has correct syntax but fails to build;
    \item \textit{Failing Test}: the test builds but fails due to wrong assertions or expected behavior;
    \item \textit{Passing Test}: the test builds and passes;
    \item \textit{Correct Test}: the test passes and covers the correct focal method;
\end{itemize}

The first four categories are mutually exclusive, that is a $tc$ can either be classified in \textit{Syntax Error} or \textit{Failed Build} and so on, while the last category (\textit{Correct Test}) represents a more stringent subset of the \textit{Passing Test} one. Specifically, we consider a test case $tc$ to be correct only if it builds properly, executes without failing, and covers the correct focal method given as input.

We report statistics for all the generated test cases and defined categories. Additionally, we report method-level statistics considering the percentage of methods successfully tested, that is, those that have at least one correct test case out of the 30 candidates.

In total, we consider 5 projects, 161 different revisions, and generate test cases for 5,278 focal methods.

\textbf{RQ$_5$: How does our approach compare to EvoSuite and GPT-3?}
The goal of this research question is to provide a preliminary quantitative and qualitative comparison between the test cases generated by our model and those generated by two alternative approaches: EvoSuite and GPT-3. We chose these two approaches as representative of two different classes of techniques: (i) evolutionary-based automated test case generation; (ii) transformer-based language models.

\subsubsection*{EvoSuite}
EvoSuite~\cite{fraser2011evosuite} is a widely known tool that automatically generates unit tests for Java software. EvoSuite uses an evolutionary algorithm to generate JUnit tests, targeting code coverage criteria. Specifically, it introduces mutants and iteratively generates assert statements to kill such mutants. During this process, EvoSuite minimizes the number of asserts while trying to maximize the number of detected mutants.

\subsubsection*{GPT-3}
Generative Pre-trained Transformer 3 (GPT-3) is an autoregressive language model introduced by OpenAI~\cite{brown2020language}. GPT-3 is a transformer decoder-only architecture having 175 billion trainable parameters. It has been pre-trained on the Common Crawl dataset~\cite{raffel2019exploring} constituting nearly a trillion words, an expanded version of the WebText~\cite{Gokaslan2019OpenWeb} dataset, two internet-based books corpora (Books1 and Books2), and English-language Wikipedia. GPT-3 has demonstrated an impressive task-agnostic few-shot performance on text generation, translation and question-answering, as well as cloze tasks. The few-shot learning assumes an extended context supplied to the model during inference as a task description, and requires no gradient updates.

%For the unit test case generation experiment with GPT-3, we have randomly drawn two focal method--unit test case method pairs (the maximum number of shots that fit into the 2048 context window allowed by the model) from the supervised training set for our target downstream task as conditioning, delimited by \texttt{``// input focal method''} and \texttt{``// output unit test case method''} strings marking the input and the output for generation.

\subsubsection*{Experiment's Design} In this experiment, we aim at assessing two main qualities of the generated test cases: (i) \textit{correctness} -- tests that accurately assert the behavior of the focal method; (ii) \textit{code coverage} -- number of lines and conditions covered by the test cases.

For this comparison we select a small but reproducible testbed using defects4j~\cite{just2014defects4j}. We rely on defects4j since it provides a reliable infrastructure to generate, compile, execute, and evaluate test cases. Specifically, we select Lang-1-f, which represents the fixed version of the first bug in the defects4j collection belonging to the project Apache Commons Lang~\cite{apachelang}. Note that these projects are not included in our pretraining or finetuning datasets. We generate unit test cases for all the public methods of the class impacted by the bug, \texttt{NumberUtils}, using our model, EvoSuite, and GPT-3. Next, we compile and execute the test cases and manually assess their correctness. Specifically, to be defined as correct, the test case needs not only to be able to execute and pass, but also must specify at least one assert that is semantically accurate w.r.t the focal method. Subsequently, we compute test coverage using defects4j (which, in turn, relies on Cobertura~\cite{cobertura}) singularly for each unit test case generated by the three approaches.

\subsubsection*{EvoSuite - Generation}
To generate test cases with EvoSuite, we use the defects4j built-in command \texttt{gen\_tests.pl -g evosuite -p Lang -v 1f}. This command invokes EvoSuite test generation on the first fixed revision of Lang, which will generate test cases for the class affected by the bug (\ie \texttt{NumberUtils}). We let EvoSuite generate test cases for 500 seconds ($\sim8$ minutes). Then, we test every unit test case generated and select the best test case for each focal method.

\subsubsection*{GPT-3 - Generation}
To generate test cases with GPT-3 we rely on few-shot learning. Specifically, we provide two examples of input focal method and corresponding test case taken from the training set, then feed one of the public methods in the \texttt{NumberUtils} class, and expect GPT-3 to answer with the corresponding test case. 

We use the OpenAI API and \texttt{davinci-msft} serving endpoint to perform inference on the model. We experiment with two different sets of prompts (\ie focal methods and test cases) from the supervised training set for our target downstream task as conditioning, varying the sampling temperature parameter from 0.1 to 0.9 with 0.1 increments (\ie the higher the temperature, the more risky or creative are the outputs). We generate ten candidate output sequences for each focal method, selecting the best test case for each focal method. Note, we fall back to one-shot learning if the examples and the current focal method exceed the maximum sequence length for GPT-3 (\ie 2048 tokens), which happened only once.

\subsubsection*{\approach - Generation}
The generation process is similar to what is described in the previous RQ$_4$, except that we select the best prediction from the top-10 candidates, rather than the top-30. 

We are aware that this represents only a small-scale preliminary evaluation, however, given the significant manual effort assessing the correctness, we believe this is an important first step. We discuss this in the threats to validity section.

\textbf{RQ$_6$: Do developers prefer \approach's test cases over EvoSuite's?}
In this research question we aim at analyzing the developer's perspective and preferences regarding test cases. In particular, we are interested in developers' view of different aspects of test cases, such as readability, understandability, and testing effectiveness.

To this aim, we designed a survey with developers where we show them a focal method under test and two alternative test cases: one generated with \approach, and the other with EvoSuite. We then pose the developers three questions, asking that they rely on their personal preferences when evaluating these factors:

\begin{itemize}[leftmargin=*]
    \item Q$_1$: Which test case is more readable and understandable?
    \item Q$_2$: Which test case is testing the method more appropriately?
    \item Q$_3$: Which test case would you prefer to have in your project?
\end{itemize}

The first two questions are designed to evaluate two different factors, namely understandability and testing effectiveness of the test cases. These questions can be answered by choosing: (i) Test Case A; (ii) Test Case B; (ii) Equally (\ie same degree of understandability and testing effectiveness). The third question is designed to break possible ties, and asks for overall preference between the two test cases (choose A or B). This will provide some clues as to whether developers prefer one factor over the other.

The survey consists of two background questions, asking about Java and JUnit experience, followed by 14 testing scenarios to review. Each scenario is formed by a focal method, and two test cases (one from \approach, the other from EvoSuite), randomly assigned with label A or B. The 14 focal methods have been selected from the experiment in RQ$_5$ and all the test cases selected are compilable and correct. We simply instruct the developer to answer the questions based on their personal preferences, without providing any clues about which test case was generated by our approach.

\section{Experimental Results}
In this section we report and discuss the results of our empirical study.

\begin{figure}[t!]
    \centering
    \vspace{-0.1cm}
    \caption{Pretraining Models - Validation Loss \\ English and Code pretraining provide positive effect}
    \vspace{-0.1cm}
    \includegraphics[width=0.47\textwidth]{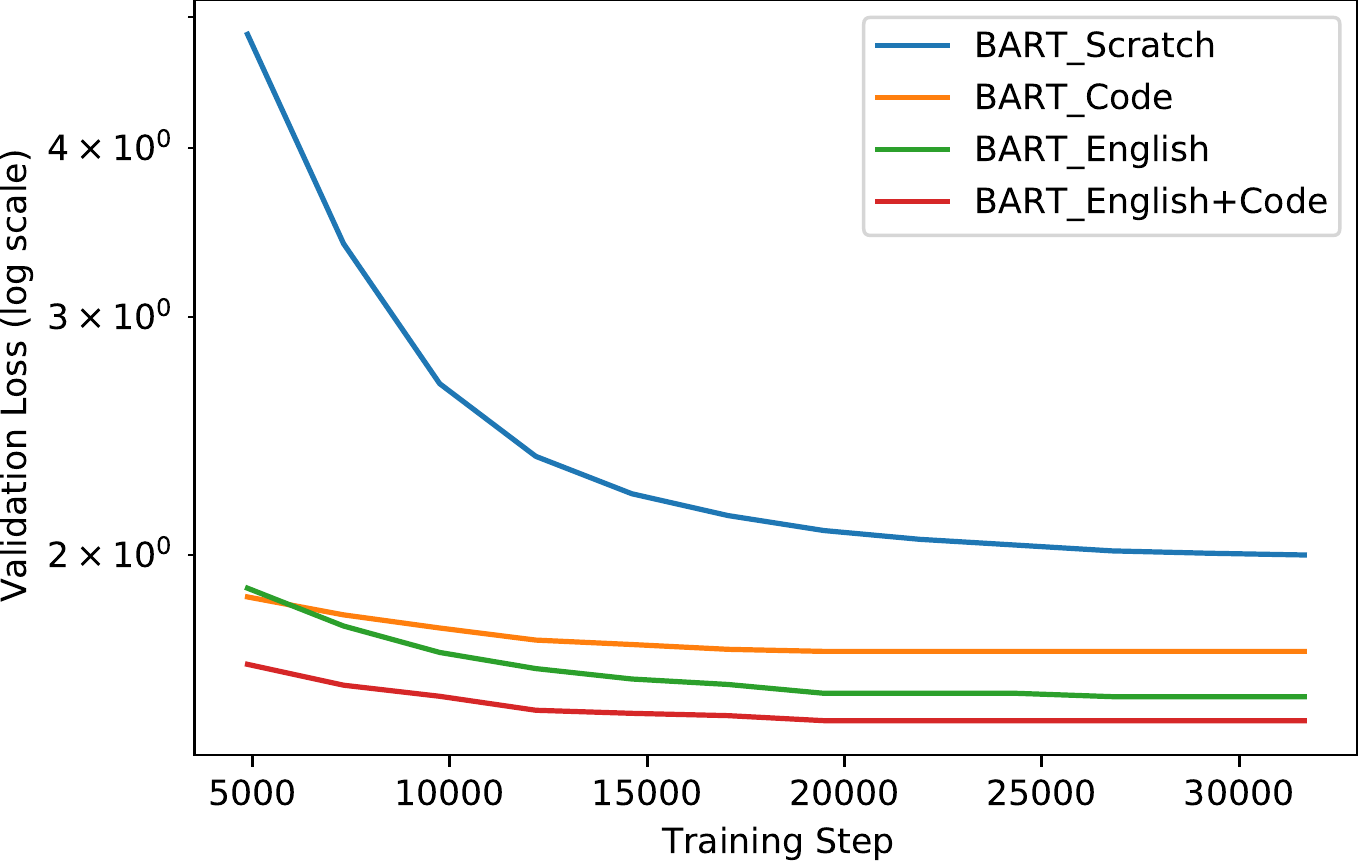}
    \vspace{-0.5cm}
    \label{fig:loss_pretrain}
\end{figure}

\textbf{RQ$_1$: Does model pretraining impact the performances of Unit Test Case Generation?}
Figure \ref{fig:loss_pretrain} shows the cross-entropy loss on the validation set during training for the four model variations. We note a substantial gap between the model without pretraining (\textit{BART\_Scratch}) compared to the models with English (\textit{BART\_English}), source code (\textit{BART\_Code}) and both (\textit{BART\_English+Code}) pretraining. Comparing the English only and the English+Code models, the additional pretraining on source code has three evident effects: (i) lower initial loss (1.89 versus 1.66); (ii) lower best loss (1.56 versus 1.51); (iii) faster convergence ($\sim$20k training steps earlier).

We conclude that English and Code pretraining are beneficial for our downstream task, thus we select the \textit{BART\_English+Code} as our starting model for the subsequent finetuning steps.

\vspace{-0.2cm}
\begin{center} 
\fbox{
\begin{minipage}[t]{0.97\linewidth}
{\bf Summary for RQ$_1$.} 
Pretraining on both English and source code has a significant positive effect on the task of generating Test Cases. The model \textit{BART\_English+Code} achieves the best validation loss.
\end{minipage}
}
\end{center}
\vspace{0.2cm}

% ---------------------------------------------------

\textbf{RQ$_2$: How does focal context impact the training for Unit Test Case Generation?} 
In this section we report the results of our experiments aiming at investigating the impact of the focal context on the test case generation task.

\subsubsection*{Ingredient Space Analysis}
\begin{figure}[h!]
    \centering
    \caption{Focal Context - Ingredient Analysis \\ Ingredients for tests are available in the focal context}
    \vspace{0.1cm}
    \includegraphics[width=0.47\textwidth]{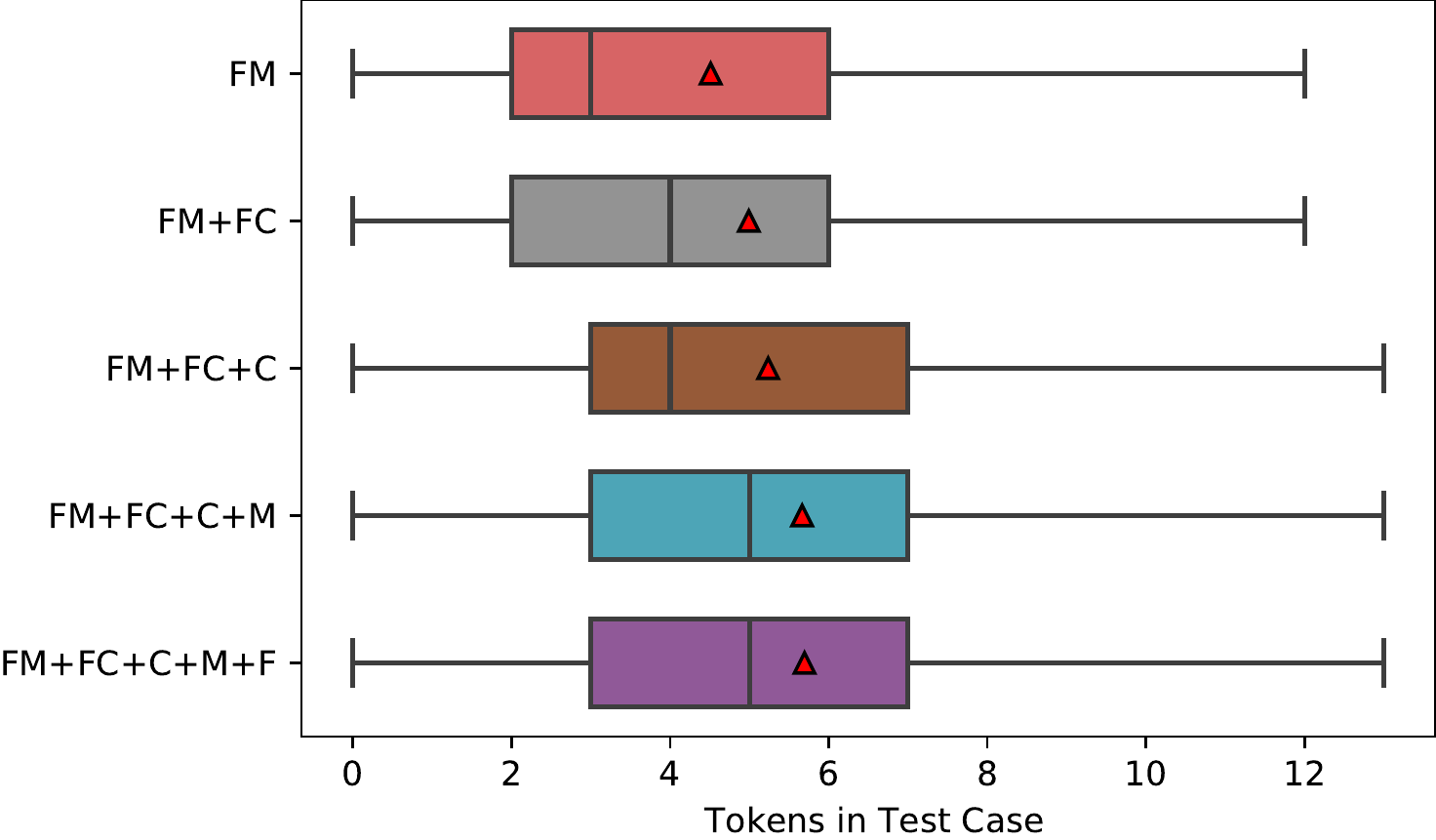}
    \vspace{-0.0cm}
    \label{fig:ingredient}
\end{figure}

Figure \ref{fig:ingredient} shows the distribution of number of tokens in the target test case that are shared with the input code representations. The distributions are represented with boxplots, where the vertical line represent the median and the red triangle the mean.

The first representation (\textit{fm}) shares 3 tokens on median and 4.15 tokens on average with the target test case, while the largest representation (\textit{fm+fc+c+m+f}) shares 5 tokens on median and 5.69 tokens on average with the corresponding test case.

From the boxplots we can notice that the focal method represents the major contribution to the test case, in terms of ingredients. The focal class name and the constructors boost significantly the shared tokens, while the subsequent additions to the focal context have diminishing returns.

This preliminary analysis confirms the intuition that additional focal context can provide useful ingredients to the model when generating test cases.

\begin{figure}[t]
    \centering
    \vspace{-0.1cm}
    \caption{Focal Context Models - Validation Loss \\ Additional focal context improves task loss}
    \vspace{-0.1cm}
    \includegraphics[width=0.47\textwidth]{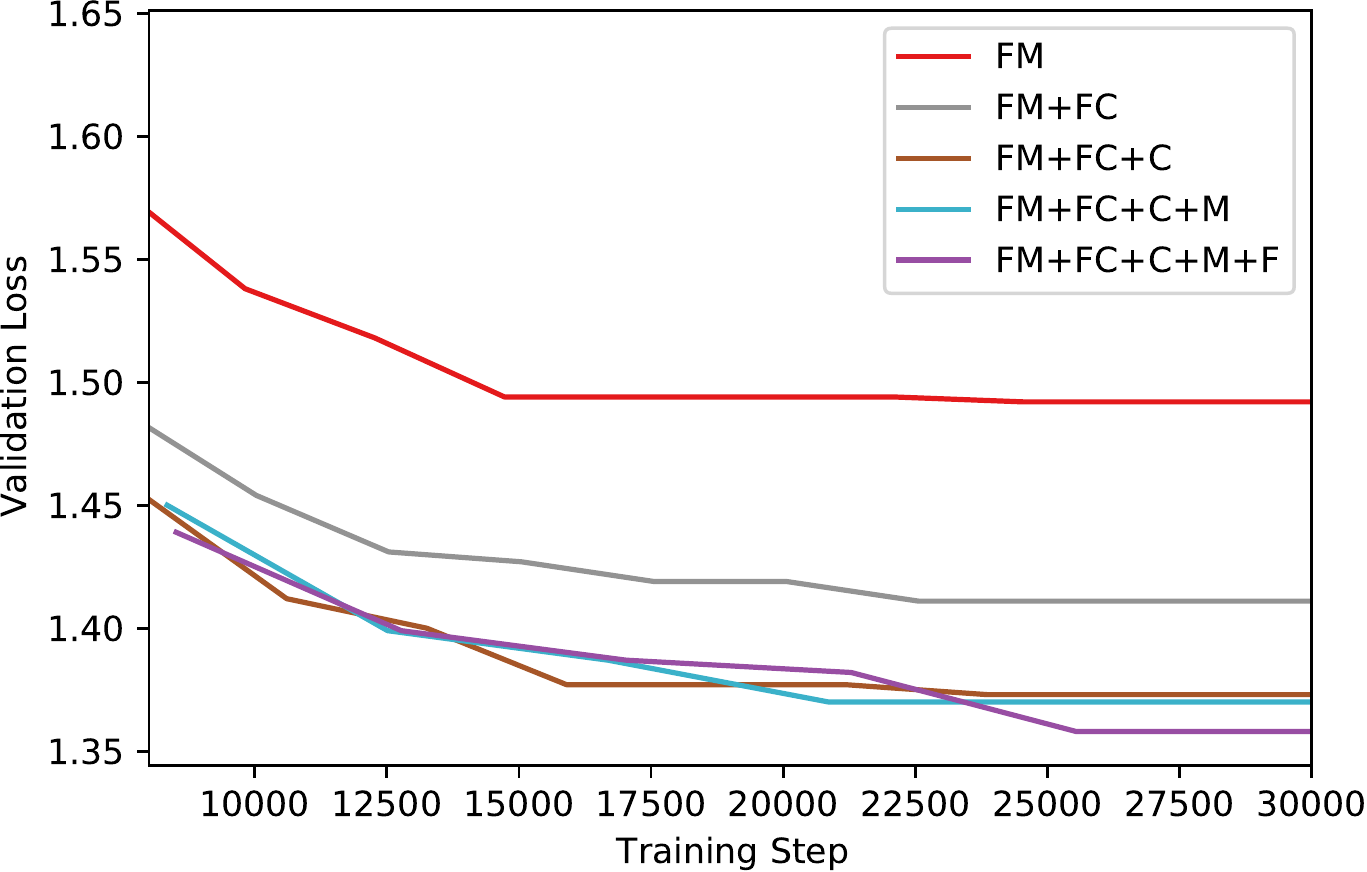}
    \vspace{-0.5cm}
    \label{fig:loss_context}
\end{figure}

\subsubsection*{Validation Loss}
Figure \ref{fig:loss_context} shows the cross-entropy loss on the validation set during training for the five focal context model variants. All the model variants have been finetuned starting from the \textit{BART\_English+Code}, which was selected as the best pretrained model in the previous research question. 

The model variant \textit{fm} depicted with a red line in Fig. \ref{fig:loss_context} corresponds to the red line in Fig. \ref{fig:loss_pretrain}, which is the \textit{BART\_English+Code} model trained with the minimal focal context (\textit{fm}).

The model variants with additional focal context show improved validation loss over the base \textit{fm} model. Specifically, the biggest delta improvement is observed when adding the focal class name (\textit{fm+fc}). This representation has only few additional tokens compared to the \textit{fm} model, however they appear to provide significant boost during training. We hypothesize that the focal class name is a strong semantic clue that can be leveraged by the model when generating tests.

The next three model variants \textit{fm+fc+c}, \textit{fm+fc+c+m}, and \textit{fm+fc+c+m+f} cluster together towards the bottom of the graph, with significant improvement over the first two variants (\textit{fm} and \textit{fm+fc}). Overall, the best performing model is the \textit{fm+fc+c+m+f}, which has the largest available focal context.

These results confirm that focal context, in addition to the focal method, provides informative tokens upon which the model can attend while generating unit test cases. The ingredient analysis complemented with the validation loss analysis corroborates the intuition that information from the focal class, such as its constructors, methods, and fields, are beneficial to the downstream task. 

We select the model \textit{BART\_English+Code} pretrained on English and code, then finetuned with the representation \textit{fm+fc+c+m+f}, as our target model for \approach.

\vspace{-0.2cm}
\begin{center} 
\fbox{
\begin{minipage}[t]{0.97\linewidth}
{\bf Summary for RQ$_2$.} 
Focal context improves the performances of the model. It provides token ingredients that can be used during the generation of unit test cases. The model \textit{fm+fc+c+m+f}, with the largest available focal context, achieves the best validation loss.
\end{minipage}
}
\end{center}
\vspace{0.5cm}

% ---------------------------------------------------

\begin{figure}[t]
    \vspace{-0.1cm}
    \centering
    \caption{Testing APIs Distribution \\ Generated tests contains similar number of testing APIs}
    \vspace{0.1cm}
    \includegraphics[width=0.47\textwidth]{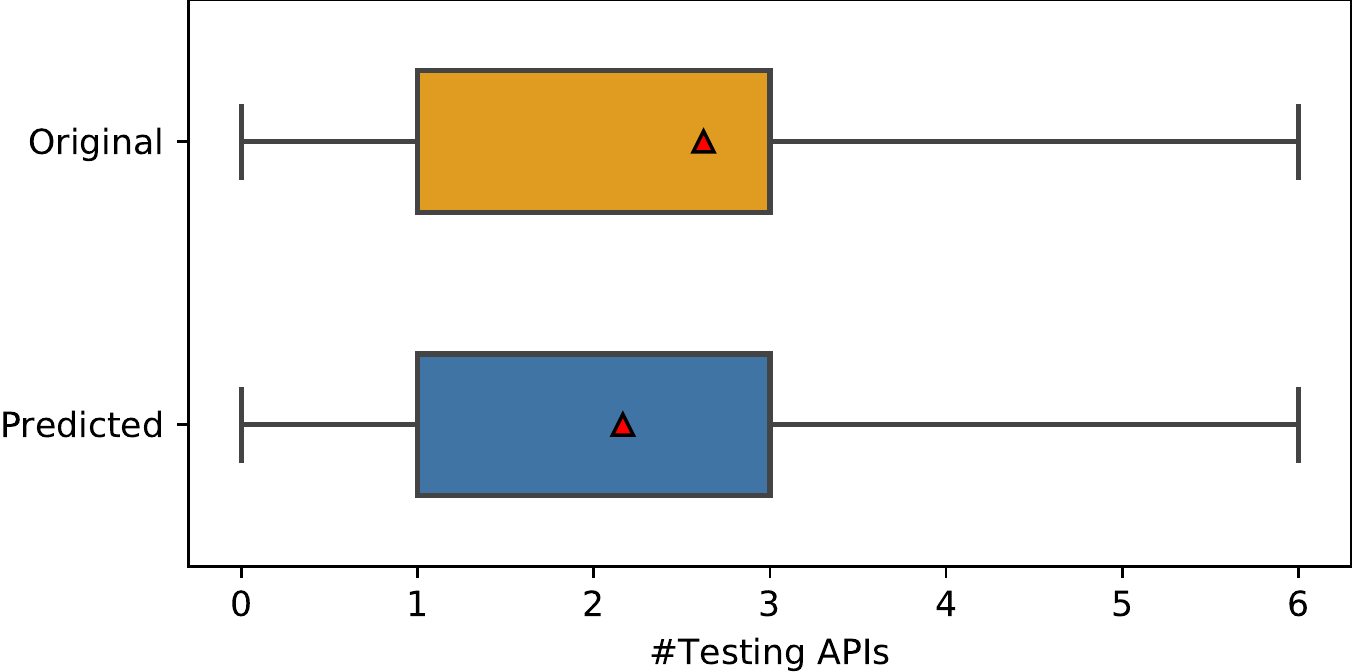}
    \vspace{-0.4cm}
    \label{fig:apis_boxplot}
\end{figure}

\vspace{0.2cm}
\textbf{RQ$_3$: What is quality of the generated Test Cases?}

\subsubsection*{Syntactic Correctness}
The model generates syntactically correct Java methods for 84\% of the top predictions in the test set. We manually investigated the reasons behind the syntactic errors for some of the predictions, and found that they were mostly due to truncated sequences when generating long test cases. We devised a simple approach that attempts to recover these predictions by deleting the last truncated statement, and adding a closing parenthesis. With this simple approach, the syntactic correctness reaches 95\%. These results show that our approach is able to generate syntactically correct Java methods in most of the cases, and with simple post-processing it achieves extremely high levels of correctness. Furthermore, an incorrect prediction could be replaced with another prediction generated by the model (on the same focal method) using beam search or sampling.

\begin{figure}[t]
    \vspace{-0.1cm}
    \centering
    \caption{Testing APIs Breakdown Distribution \\ Generated tests contains similar API distribution to original}
    \vspace{0.1cm}
    \includegraphics[width=0.47\textwidth]{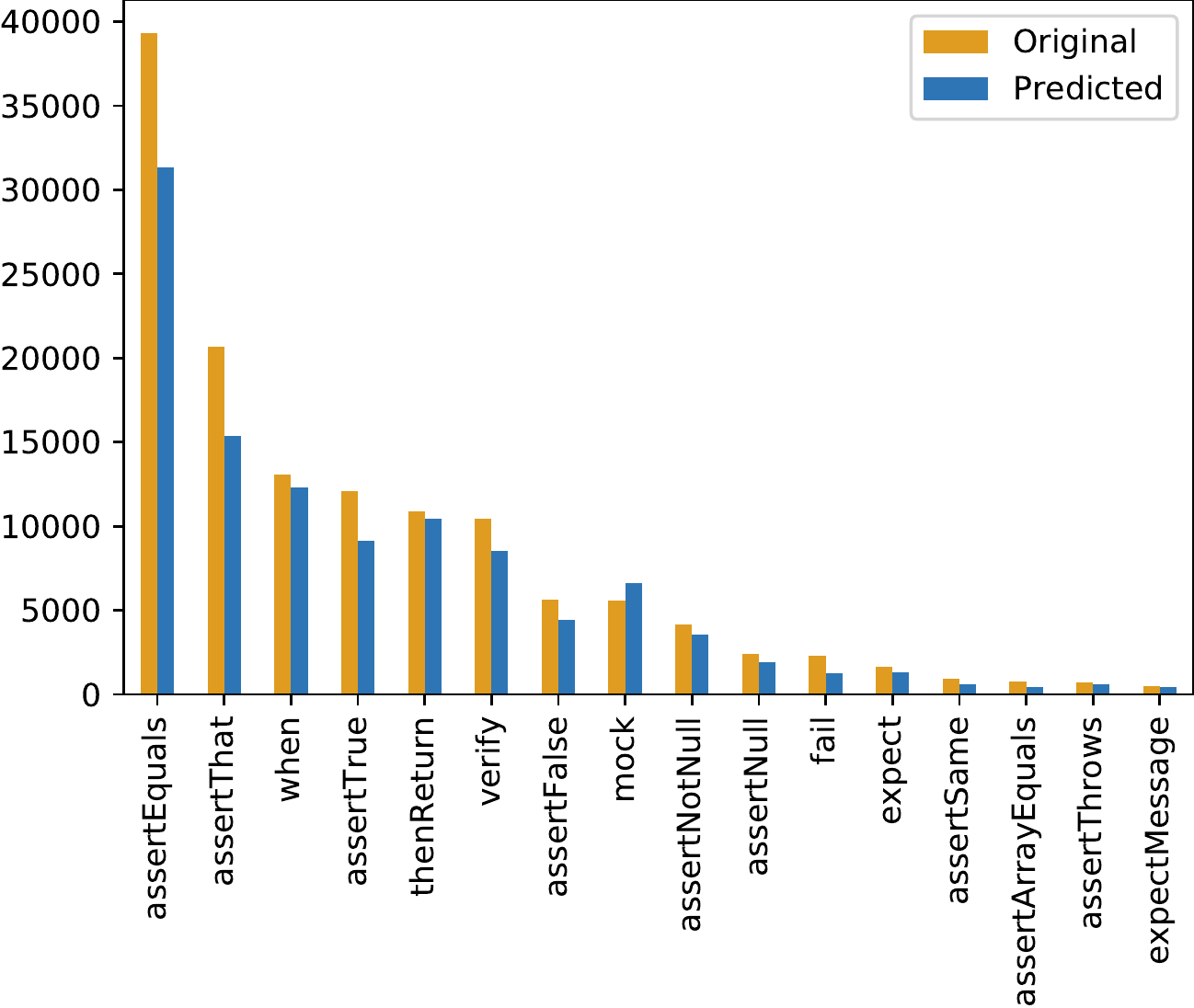}
    \vspace{-0.5cm}
    \label{fig:apis_barplot}
\end{figure}

\subsubsection*{Testing APIs}
The model generates methods that declare the \texttt{@Test} annotation in 99.99\% of the cases, correctly learning the JUnit standard for test cases. Furthermore, 94.9\% of the generated test cases invoke the correct focal method which is supposed to test.

Figure \ref{fig:apis_boxplot} shows the distribution of testing API calls within each test cases in the test set, both for the original test cases and for the predictions of the model. From the boxplot we can notice that the two distributions have the same quartiles with, on median, one testing API call in each test case. Note that outliers are not reported in this figure. The mean (shown as a red triangle) indicates that the original test cases tend to contain slightly more testing APIs compared to the ones generated by the model.

Figure \ref{fig:apis_barplot} shows the breakdown distribution of the top-16 testing API found in the test set. These include JUnit APIs such as \texttt{assertEquals} and Mockito APIs such as \texttt{mock} and \texttt{verify}. The plot clearly shows that the generated test cases invoke a variety of different testing APIs, closely following the distribution of the original test cases. However, we do observe a gap between the number of APIs in the original and predicted test cases. In our future work we plan to incorporate techniques to augment the number of assert statements in the test cases.

%For improved visibility we report only the top-16 testing APIs, however we analyzed a total of 20 testing APIs, finding that the models can generate all of them, with similar distribution to the original test cases. All the results are available on our online appendix.

We conclude this research question with qualitative examples of test cases generated by the model. Figure \ref{fig:example_generalizability} shows the focal method \texttt{review} belonging to the class \texttt{CSConference}. This example was written by one of the authors with the goal of demonstrating the generalizability of our model on \textit{novel} and \textit{unusual} input methods (\ie not belonging to the training distribution). Figure \ref{fig:example_generalizability} also shows three examples of generated test cases selected from the top-10 predictions of the model. In the test case \#1 the model creates a new instance of \texttt{Submission}, followed by an assert statement that contains the focal method invocation. The model generates a correct method invocation passing the correct number of arguments and types. The assert statement interestingly inspects the reason of the \texttt{Response} by matching the same string used within the focal method. In the test case \#2 the model first creates a realistic instance of a paper submission, then it invokes the focal method saving its response, and finally asserts that the response is of type \texttt{Acceptance}. Finally, in test case \#3 the model relies on mocking APIs from Mockito to mock the \texttt{Submission} class, as well as using assert statements on the response. These examples showcase the ability of our model in generating realistic test cases, which resemble test cases written by developers.

\vspace{-0.2cm}
\begin{center} 
\fbox{
\begin{minipage}[t]{0.97\linewidth}
{\bf Summary for RQ$_3$.} 
Our approach generates syntactically correct test cases that conform to the Unit Test Case standards. The generated test cases invoke a variety of Testing APIs, following a similar distribution to the original test cases. 
\end{minipage}
}
\end{center}

% ---------- Code Example ----------------
\begin{figure}[t]
\vspace{-0.2cm}
    \centering
\begin{adjustbox}{width=0.5\textwidth}
\footnotesize
\begin{tabular}{c}
\toprule
  Focal Method \\
\midrule
\begin{minipage}[t]{0.45\textwidth}
\begin{minted}{java}
public class CSConference { 
public static Response review(Submission paper, List<Reviewer> revs) {
    if(paper.citations < 50){
        return new Rejection("Discuss related works");
    }
    if(!paper.hasComparison()){
        return new Rejection("Compare with state of the art");
    }
    return new Acceptance("Congrats!");
}}
\end{minted}
\end{minipage}\\
\midrule
Model Generated Test Case \#1 - Asserts\\
\midrule
\begin{minipage}[t]{0.45\textwidth}
\begin{minted}{java}
 @Test public void conference() {
    Submission paper = new Submission();
    assertTrue(CSConference.review(paper, new ArrayList<Reviewer>()).getReason()
        .equalsIgnoreCase("Discuss related works"));
 }
\end{minted}
\end{minipage} \\
\midrule
Model Generated Test Case \#2 - Asserts\\
\midrule
\begin{minipage}[t]{0.45\textwidth}
\begin{minted}{java}
 @Test public void mentionedStates() {
    Submission paper = new Submission("Yang et al. 2012", 
        "Elements of Information Sciences", 61, 3);
    Response response = CSConference.review(paper, new ArrayList<Reviewer>());
    assertTrue(response instanceof Acceptance);
}
\end{minted}
\end{minipage} \\
\midrule
Model Generated Test Case \#3 - Mockito \& Asserts\\
\midrule
\begin{minipage}[t]{0.45\textwidth}
\begin{minted}[escapeinside=||]{java}
@Test public void reference() {
    Submission paper = mock(Submission.class);
    when(paper.getNumberOfCitations()).thenReturn(3);
    when(paper.citations()).thenReturn(60);
    when(paper.hasComparison()).thenReturn(true);
    Response response = Conference.review(paper, new ArrayList<Reviewer>());
    assertEquals(new Acceptance("Congrats!"), response);
    when(paper.citations()).thenReturn(99);
    response = Conference.review(paper, new ArrayList<Reviewer>());
    assertEquals(new Rejection("Discuss related works"), response);
}
\end{minted}
\end{minipage} \\
\bottomrule
\end{tabular}
\end{adjustbox}
%\medskip
\vspace{-0.2cm}
\caption{Examples of Generated Test Cases}
\vspace{-0.4cm}
\label{fig:example_generalizability}
\end{figure}
% ---------- END Code Example ----------------

\begin{table*}[t]
    \centering
	\vspace{-0.4cm}
	\caption{Defects4j Results -- Test cases generated by \approach are classified in five categories based on syntax correctness, compilability, test execution, and coverage. Overall, 16.21\% of the generated tests are correct (\ie compile, pass, and cover the correct focal method), with \approach being able to correctly tests 43.75\% of the focal methods.}
	\label{tab:defects4j-results}
	\resizebox{1\linewidth}{!}{
\begin{tabular}{lrr|rrrrrr}
\toprule
\multirow{ 2}{*}{Project} & \multicolumn{2}{c}{Focal Methods} & \multicolumn{5}{c}{Test Cases} \\
& Tested & Total & Correct & Passing & Failing & Build Error & Syntax Error & Total\\
\midrule

Lang    & 1,545 (56.97\%) & 2,712 & 18,997 (23.35\%) & 25,563 (31.42\%) &  26,515 (32.58\%) & 19,450 (23.90\%) & 9,832 (12.08\%) & 81,360\\
Chart   & 425 (32.00\%) & 1,328 & 3,443 (8.64\%) & 4,663 (11.70\%) & 6567 (16.48\%) & 28,318 (71.07\%) & 292 (0.73\%) & 39,840 \\
Cli     & 190 (29.46\%) & 645 & 2,142 (11.07\%) & 2,399 (12.39\%) & 5,492 (28.38\%) & 10,545 (54.49\%) & 914 (4.72\%) & 19,350 \\
Csv     & 128 (34.31\%) & 373 & 912 (8.15\%) & 1,005 (8.98\%) & 2,160 (19.30\%) & 5,370 (47.99\%) & 2,655 (23.72\%) & 11,190 \\ 
Gson    & 21 (9.54\%) & 220 & 186 (2.80\%) & 186 (2.80\%) & 1,573 (23.71\%) & 3,497 (52.71\%) & 1,344 (20.26\%) & 6,634 \\
\midrule
Total   & \textbf{2,309 (43.75\%)} & 5,278 & \textbf{25,680 (16.21\%)} & 33,816 (21.35\%) & 42,307 (26.71\%) & 67,180 (42.41\%) & 15,037 (9.49\%) & 158,374 \\
\bottomrule
\end{tabular}}
\vspace{-0.1cm}
\end{table*}

\textbf{RQ$_4$: Can \approach generate Test Cases for Defects4j projects?}

In this section we report the results of using \approach for generating unit test cases for five defects4j projects. For each focal method, we allow \approach to generate 30 candidate test cases using beam search, and evaluate each candidate based on syntax correctness, compilability, execution, coverage and correctness.

Table \ref{tab:defects4j-results} reports the results of our experiments, where the left side of the table provides focal method-level statistics, while the right side test case-level statistics.

We begin by discussing the Test Case statistics from right to left. In our experiments \approach generated a total of 158k test cases for 5,278 focal methods. 

The syntactically incorrect test cases account for 9.49\% of the total generated tests. While this represents an opportunity and future direction for improving our model, in practice, these candidates could be easily and quickly discarded using a syntax checker.

A significant portion of the generated tests (42.41\%) are syntactically correct but fail to build. In our manual investigation, we found that these build errors are often due to incorrect usage of classes and methods outside the focal class. For example, a test case may need to instantiate an object of a different class that is used by the focal method, and an incorrect usage of the object (\eg wrong method name or parameter) may lead to a build error. 

Failing tests, which are compilable but fail during their execution, represent 26.71\% of the generated tests. These tests usually fail for incorrect assertions or wrong expected behavior (\eg the test expects an exception which is not raised).

Passing tests account for 21.35\% of the tests generated by \approach. These tests are syntactically correct, compilable, and execute without failing.

Finally, when analyzing the coverage information of the passing test cases, we classify 16.21\% of all the generated test cases, $\sim$25K tests, as \textit{correct}. These test cases are a subset of the passing tests which cover the correct focal method given as input. Note that the remaining passing test cases that are not covering the focal method, could potentially still be used to test other parts of the project under test.

Considering the focal method-level statistics, \approach was able to generate at least one correct test case for 43.75\% of all the focal methods, for a total of $\sim$2k different methods. We believe that this percentage could be increase by allowing the model to generate additional test cases over the first 30 candidates.

Overall, the results of our experiments demonstrate that \approach is able to correctly test a large number of different focal methods belonging to a diverse set of projects.

While a 16\% correct rate for candidate tests could be perceived as an underwhelming result, it is worth noting that we are disclosing and analyzing every single attempt by our model. Common automated test generation approaches often create many \textit{internal} candidates that are mutated, analyzed, and discarded before the correct ones are presented to the user. For example, EvoSuite can generate a large offspring set, where descendants are mutated, evaluated, and discarded during the evolution.

In a deployment setting, \approach could be used to generate a large set of candidates which are then analyzed and filtered before being introduce in the project under test.

We publicly release all 25K correct test cases generated by \approach in this experiment \cite{athenatestwebsite}. Test cases and associated metadata (\eg coverage files) are available for browsing and downloading at our dedicated website. \footnote{\url{https://athenatestdemowebsite.azurewebsites.net}}

\vspace{-0.2cm}
\begin{center} 
\fbox{
\begin{minipage}[t]{0.97\linewidth}
{\bf Summary for RQ$_4$.} 
\approach is able to generate correct test cases for different defects4j projects. When generating up to 30 candidates, \approach was able to correctly tests 43\% of all the focal methods, with 16\% of the candidate tests being correct.
\end{minipage}
}
\end{center}

% ------------------------------------------------------

\textbf{RQ$_5$: How does our approach compare to EvoSuite and GPT-3?}

Table \ref{tab:coverage} reports the results of our test coverage analysis comparing EvoSuite, GPT-3, and \approach on the class \texttt{NumberUtils} of Lang-1-f. The table reports the absolute (and percentage) line and condition coverage at class-level, for each of the 18 unique public methods in the class (without considering overloading), marking in bold the best coverage value. From the results in Table \ref{tab:coverage} we notice: (i) EvoSuite was able to successfully test all the methods; (ii) GPT-3 correctly tested only 6 out of 18 methods; (iii) \approach generated correct test cases for all the methods, while achieving the best coverage in most cases.

For GPT-3 we explored several sampling temperatures, and settled on the 0.5 value which appeared to provide good diversity of the samples while still generating realistic code. We found that, in most of the cases where GPT-3 was not able to generate a correct test case, it generated code that only invoked the focal method without correctly asserting its behavior. However, in those 6 cases reported in the table, we found the test cases to be correct and readable code, and sometimes also obtaining the best coverage. While GPT-3 achieved the lowest overall performances of the three, we would consider this still a positive result for GPT-3, given the fact that it was not finetuned on test case generation.

Regarding our approach, \approach was able to generate correct test cases for all the focal methods. Overall, the results indicate that \approach is able to generate correct test cases with adequate test coverage, often achieving better coverage than EvoSuite.

\begin{table*}[t]
	\vspace{-0.4cm}
	\caption{Test Coverage Analysis -- Test cases generated by EvoSuite, GPT-3, and \approach are executed and their coverage is analyzed in terms of line and condition covered. \approach has a comparable coverage w.r.t. EvoSuite.}
	\label{tab:coverage}
	\centering
	\resizebox{0.9\linewidth}{!}{
\begin{tabular}{lcccccc}
\toprule
\multirow{ 2}{*}{Focal Method} & \multicolumn{2}{c}{EvoSuite} & \multicolumn{2}{c}{GPT-3} & \multicolumn{2}{c}{\approach}\\
& Lines & Conditions & Lines & Conditions & Lines & Conditions\\
\midrule
\texttt{toInt(String, int)} & 21 (5.6\%) & 1 (0.3\%) & - & - & \textbf{23 (6.1\%)} & \textbf{2 (0.6\%)}\\
\texttt{toLong(String, long)} & \textbf{20 (5.3\%)} & \textbf{1 (0.3\%)} & - & - & \textbf{20 (5.3\%)} & \textbf{1 (0.3\%)}\\
\texttt{toFloat(String, float)} & 20 (5.3\%) & 1 (0.3\%) & - & - & \textbf{22 (5.9\%)} & 1 (0.3\%)\\
\texttt{toDouble(String, double)} & \textbf{20 (5.3\%)} & \textbf{1 (0.3\%)} & - & - & \textbf{20 (5.3\%)} & \textbf{1 (0.3\%)}\\
\texttt{toByte(String, byte)} & 20 (5.3\%) & 1 (0.3\%) & - & - & \textbf{23 (6.1\%)} & \textbf{2 (0.6\%)}\\
\texttt{toShort(String, short)} & 20 (5.3\%) & \textbf{1 (0.3\%)} & - & - & \textbf{22 (5.9\%)} & \textbf{1 (0.3\%)}\\
\texttt{createFloat(String)} & 20 (5.3\%) & 1 (0.3\%) & - & - & \textbf{21 (5.6\%)} & \textbf{2 (0.6\%)}\\
\texttt{createDouble(String)} & 20 (5.3\%) & 1 (0.3\%) & - & - & \textbf{21 (5.6\%)} & \textbf{2 (0.6\%)}\\
\texttt{createInteger(String)} & \textbf{20 (5.3\%)} & \textbf{1 (0.3\%)} & - & - & 21 (5.5\%) & 2 (0.6\%)\\
\texttt{createLong(String)} & 20 (5.3\%) & 1 (0.3\%) & 20 (5.3\%) & 1 (0.3\%) & \textbf{21 (5.6\%)} & \textbf{2 (0.6\%)}\\
\texttt{createBigInteger(String)} & 28 (7.5\%) & \textbf{8 (2.4\%)} & \textbf{30 (8.7\%)} & 7 (2.1\%) & 20 (5.3\%) & 1 (0.3\%)\\
\texttt{createBigDecimal(String)} & \textbf{22 (5.9\%)} & \textbf{3 (0.9\%)} & - & - & \textbf{22 (5.9\%)} & \textbf{3 (0.9\%)}\\
\texttt{min(long[])} & \textbf{27 (7.2\%)} & \textbf{6 (1.8\%)} & 26 (6.9\%) & 5 (1.5\%) & 22 (5.9\%) & 2 (0.6\%)\\
\texttt{min(int, int, int)} & 22 (5.9\%) & \textbf{2 (0.6\%)} & \textbf{23 (6.1\%)} & \textbf{2 (0.6\%)} & 22 (5.9\%) & \textbf{2 (0.6\%)}\\
\texttt{max(float[])} & \textbf{28 (7.5\%)} & \textbf{7 (2.1\%)} &  - & - & 22 (5.8\%) & 2 (0.6\%)\\
\texttt{max(byte, byte, byte)} & \textbf{23 (6.1\%)} & \textbf{2 (0.6\%)} & 21 (5.6\%) & \textbf{2 (0.6\%)} & 22 (5.9\%) & \textbf{2 (0.6\%)}\\
\texttt{isDigits(String)} & 20 (5.3\%) & 1 (0.3\%) & \textbf{23 (6.1\%)} & \textbf{5 (1.5\%)} & \textbf{23 (6.1\%)} & \textbf{5 (1.5\%)}\\
\texttt{isNumber(String)} & 44 (11.7\%) & 29 (8.6\%) & - & - & \textbf{51 (13.6\%)} & \textbf{41 (12.1\%)}\\
\bottomrule
\end{tabular}}
\vspace{-0.1cm}
\end{table*}

We now provide a qualitative comparison of the test cases generated by the three approaches. Figure \ref{fig:create-float} shows the generated test cases for the focal method \texttt{createFloat}. EvoSuite creates a test case that assert that the return value of the method is null, when providing a null string as input, covering the first condition in the focal method. GPT-3 creates a test case that simply invokes the focal method multiple times (limited in the figure), with correct arguments, but without asserting the correct behaviour of the method. \approach generated a test case that checks (i) the focal method correctly creates a the float 1.2; (ii) the focal method returns null on a null string. Specifically, it covers both conditions of the focal method. We can also notice that the generated test case has a very idiomatic name \texttt{testCreateFloat} (similar to GPT-3), compared to EvoSuite's \texttt{test044}.

\begin{figure}[t]
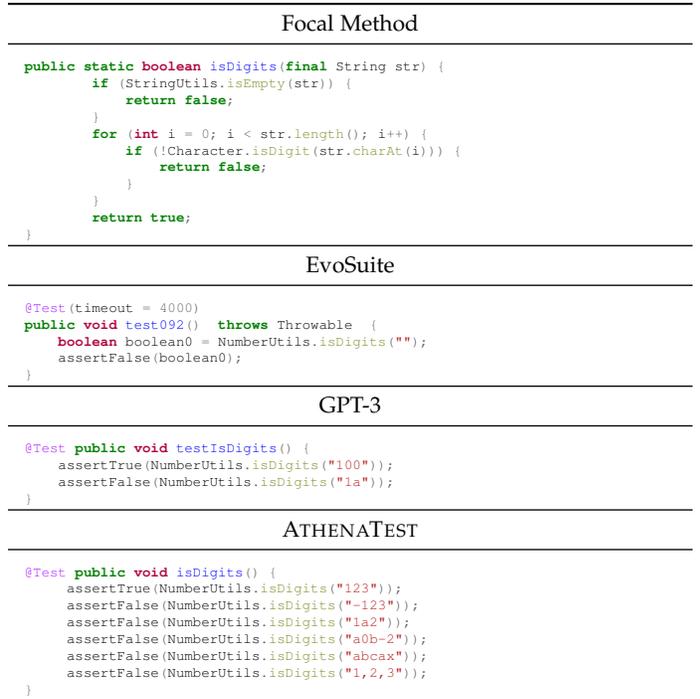

\vspace{-0.3cm}
    \centering
\begin{adjustbox}{width=0.5\textwidth}
\footnotesize
\begin{tabular}{c}
\toprule
  Focal Method \\
\midrule
\begin{minipage}[t]{0.45\textwidth}
\begin{minted}{java}
public static boolean isDigits(final String str) {
        if (StringUtils.isEmpty(str)) {
            return false;
        }
        for (int i = 0; i < str.length(); i++) {
            if (!Character.isDigit(str.charAt(i))) {
                return false;
            }
        }
        return true;
}
\end{minted}
\end{minipage}\\
\midrule
EvoSuite\\
\midrule
\begin{minipage}[t]{0.45\textwidth}
\begin{minted}[escapeinside=||]{java}
@Test(timeout = 4000)
public void test092()  throws Throwable  {
    boolean boolean0 = NumberUtils.isDigits("");
    assertFalse(boolean0);
}
\end{minted}
\end{minipage} \\
\midrule
GPT-3\\
\midrule
\begin{minipage}[t]{0.45\textwidth}
\begin{minted}[escapeinside=||]{java}
@Test public void testIsDigits() { 
    assertTrue(NumberUtils.isDigits("100")); 
    assertFalse(NumberUtils.isDigits("1a")); 
}
\end{minted}
\end{minipage} \\
\midrule
\approach\\
\midrule
\begin{minipage}[t]{0.45\textwidth}
\begin{minted}{java}
@Test public void isDigits() {
     assertTrue(NumberUtils.isDigits("123"));
     assertFalse(NumberUtils.isDigits("-123"));
     assertFalse(NumberUtils.isDigits("1a2"));
     assertFalse(NumberUtils.isDigits("a0b-2"));
     assertFalse(NumberUtils.isDigits("abcax"));
     assertFalse(NumberUtils.isDigits("1,2,3"));
}
\end{minted}
\end{minipage} \\
\bottomrule
\end{tabular}
\end{adjustbox}
%\medskip
\vspace{-0.0cm}
\caption{Test Cases Generated for \texttt{isDigits}}
\vspace{-0.3cm}
\label{fig:is-digits}
\end{figure}

Figure \ref{fig:is-digits} shows the test cases for the focal method \texttt{isDigits}. EvoSuite's test case checks whether the empty string is correctly identified as not being a numerical digit. GPT-3 accurately asserts the behavior of the method by testing a string containing only digits (\eg "100") and one that contains a non-digit character (\eg "1a"). Our approach, generates a test case which tests six different strings, one of which only contains digits, while the other five contains also non-digit characters, for example a negative number (\eg "-123") as well as comma-separated numbers (\eg "1,2,3"). Interestingly, in this case, combining EvoSuite and \approach would lead to better overall code coverage.

Compared to EvoSuite, our approach is able to generate correct test cases which obtain comparable (if not better) test coverage for most of the focal methods. \approach's test cases appear to be more readable and similar to developers-written code w.r.t. those generated by EvoSuite. Our approach outperforms GPT-3 in terms of number of correct test cases generated. While GPT-3 was not finetuned on the test case generation task, it is a substantially larger model (175 billion parameters) compared to our transformer-based model (400 million parameters).

%CreateFloat
% ---------- Code Example ----------------
\begin{figure}[t]
\vspace{-0.3cm}
    \centering
\begin{adjustbox}{width=0.5\textwidth}
\footnotesize
\begin{tabular}{c}
\toprule
  Focal Method \\
\midrule
\begin{minipage}[t]{0.45\textwidth}
\begin{minted}{java}
public static Float createFloat(final String str) {
        if (str == null) {
            return null;
        }
        return Float.valueOf(str);
}
\end{minted}
\end{minipage}\\
\midrule
EvoSuite\\
\midrule
\begin{minipage}[t]{0.45\textwidth}
\begin{minted}[escapeinside=||]{java}
@Test(timeout = 4000)
public void test044()  throws Throwable  {
   Float float0 = NumberUtils.createFloat((String) null);
   assertNull(float0);
}
\end{minted}
\end{minipage} \\
\midrule
GPT-3\\
\midrule
\begin{minipage}[t]{0.45\textwidth}
\begin{minted}[escapeinside=||]{java}
@Test public void testCreateFloat() {
    NumberUtils.createFloat("1.0");
    NumberUtils.createFloat("1.0f");
    NumberUtils.createFloat("1.0E+1");
    NumberUtils.createFloat("1.0E-1");
    ...
\end{minted}
\end{minipage} \\
\midrule
\approach\\
\midrule
\begin{minipage}[t]{0.45\textwidth}
\begin{minted}{java}
@Test public void testCreateFloat() {
     assertEquals(new Float(1.2f), NumberUtils.createFloat("1.2"));
     assertNull(NumberUtils.createFloat(null));
 }
\end{minted}
\end{minipage} \\
\bottomrule
\end{tabular}
\end{adjustbox}
%\medskip
\vspace{-0.2cm}
\caption{Test Cases Generated for \texttt{createFloat}}
\vspace{-0.7cm}
\label{fig:create-float}
\end{figure}
% ---------- END Code Example ----------------

\begin{center} 
\fbox{
\begin{minipage}[t]{0.97\linewidth}
{\bf Summary for RQ$_3$.} 
Our approach generates test cases that accurately test the focal methods and obtain comparable test coverage w.r.t. EvoSuite, as well as outperforming GPT-3. These test cases appear to be similar to developer-written test cases with readable and understandable code.
\end{minipage}
}
\end{center}

% ---------- Code Example ----------------
% ---------- END Code Example ----------------

% ------------------------------------------------------

\textbf{RQ$_6$: Do developers prefer \approach's test cases over EvoSuite's?}

We received responses from 12 Microsoft developers, none of them involved in this work. All the developers had Java experience (4 with one year or less, 7 with 1-3 years, 1 with 4 or more years). Eight of them claimed to have JUnit experience. 

\begin{figure}
    \centering
    \vspace{-0.2cm}
    \includegraphics[width=0.47\textwidth]{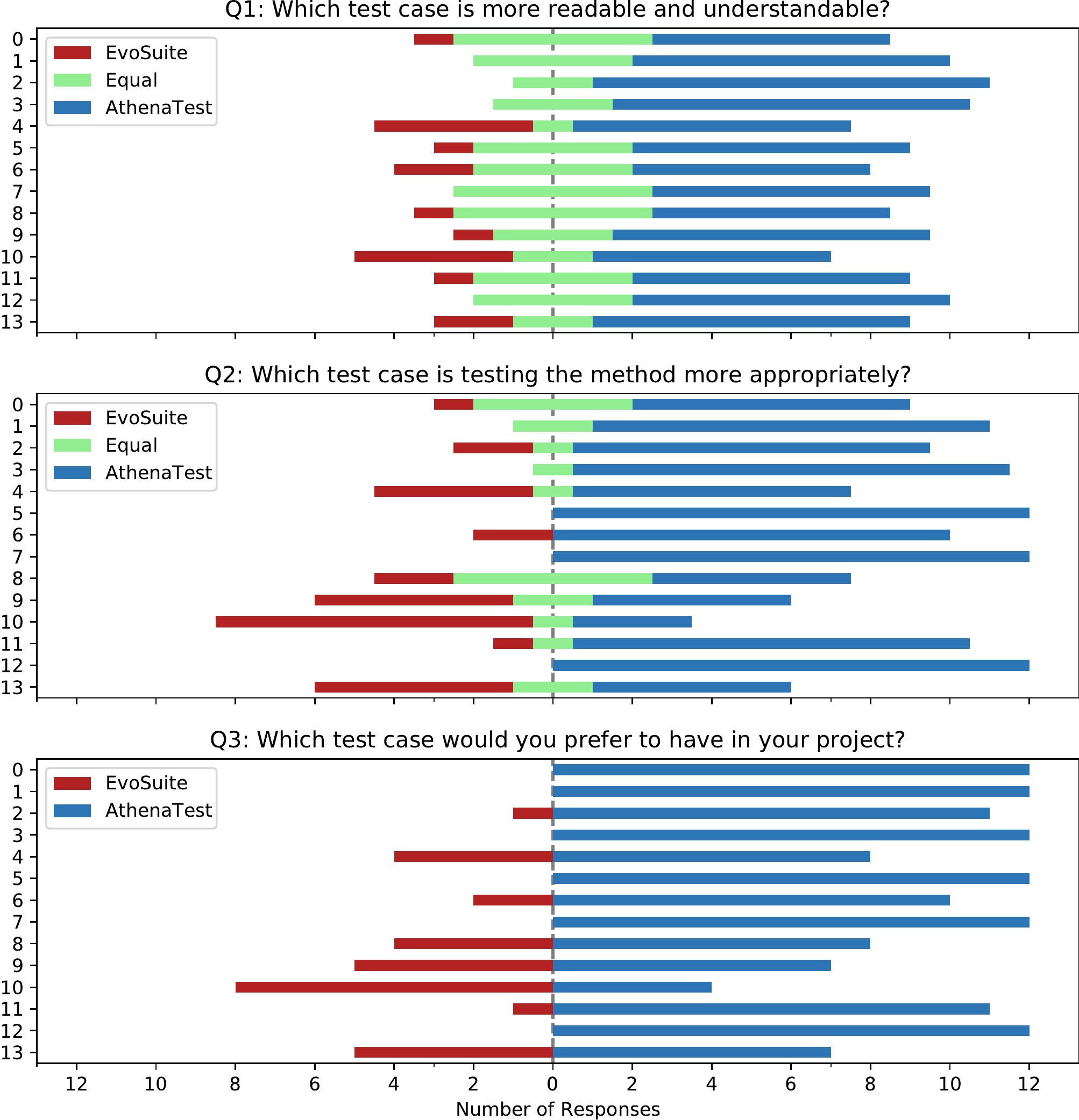}
    \vspace{-0.2cm}
    \caption{Survey results with professional developers}
    \vspace{-0.2cm}
    \label{fig:survey}
\end{figure}

Figure \ref{fig:survey} reports the answers to the three survey questions in a likert-style plot, where the y-axis represents the testing scenario instance, and the x-axis the number of responses for EvoSuite (in red, towards left), for \approach (in blue, towards right), and neutral answer (middle green).  

Regarding Q$_1$, we found that 61\% of the responses favored \approach's test cases in terms of readability and understandability, while in 29\% of the cases the developers thought both test cases were equally readable, and only in 10\% of the cases they preferred EvoSuite's. 

For Q$_2$, 70\% of the responses selected \approach's test cases as testing the focal method more appropriately than EvoSuite's counterpart. In 12\% of the cases they were deemed as equally appropriate, and only in 18\% the developers preferred EvoSuite's test case.

Finally in Q$_3$, when asked to choose which test case they preferred overall, they overwhelmingly elected \approach's test cases, in 82\% of the cases, and only 18\% EvoSuite.

Interestingly, we found that in 12 instances ($\sim$7\%), developers picked one test case in Q$_1$ and the other test case in Q$_2$. A deep dive in these cases revealed that developers mostly preferred \approach test cases in terms of readability, but EvoSuite in terms of testing effectiveness.

\vspace{-0.2cm}
\begin{center} 
\fbox{
\begin{minipage}[t]{0.97\linewidth}
{\bf Summary for RQ$_4$.} 
Developers prefer test cases generated by \approach over those generated by EvoSuite, in terms of readability, understandability, and testing effectiveness.
\end{minipage}
}
\end{center}

\section{Discussion \& Future Work}
%We trained a large transformer model to generate accurate test cases, by learning from thousands of real-world, developer-written test cases. We design our solution as a translation task, where the model is trained to translate a focal method into a corresponding test case. 
Our preliminary evaluation shows encouraging results in many different aspects. Our approach is able to generate syntactically correct test cases that conform to the test case standards and invoke a variety of testing APIs. While further analyses should be performed, this preliminary evaluation shows that the generated test cases appear to be (i) \textit{realistic} -- similar to developer-written test cases; (ii) \textit{accurate} -- correctly asserting the expected behavior of a focal method; (iii) \textit{human-readable} --  readable and understandable code, with good variable and method names.

We believe this work represents a stepping stone towards a new category of automated test case generation tools, shifting away from coverage-guided approaches towards models that aim at code understanding. These learning approaches have the potential of generating natural test cases that better integrate with the existing code base, and do not appear like \textit{machine-written} code.

During our manual investigation of the generated test cases, we also observed several weaknesses and pitfalls of the model, which we will discuss in this section. These weaknesses serve us as inspiration for future work, with the goal of improving our model.

\subsection{Project-Level Context}
When providing focal context limited to the focal method and class, the model is forced to perform a series of reasonable guesses on the composition of other classes and methods outside the scope of the focal class. For example, if the focal method takes as argument an object of a specific \textit{Class}, the model doesn't currently have knowledge about the behavior and available methods of the \textit{Class}. In those instances, the model relies on the past pretraining (on a large amount of source code) to infer the behavior of such classes.

We plan to incorporate project-level context, pertinent with the given focal method and class, in our input representation to the model. A static analysis tool could be used to collect information about the classes involved in the focal method (\eg instantiated, invoked or passed as argument), and a skeleton of such classes summarizing the APIs could be used to augment the focal context input.

Furthermore, semi-supervised pretraining on the projects where the model will be used to generate test cases could help the model to familiarize with the code base and be more accurate when generating statements and method calls.

\subsection{Testing Frameworks}
Numerous testing frameworks are available for Java developers which aim at supporting domain-specific applications or different testing scenarios and methodologies. Our current approach does not take into consideration the specific testing framework used by the developer, and thus could propose a test case using a different testing API which is not being used in the current project.

%Domain-specific testing frameworks include Selenium, which focuses on automating web-based application testing, and REST Assured which supports integration testing for REST APIs. Mockito and EasyMock are unit testing framework that allow mocking of objects which facilitate testing.

In our future work we plan to train our model to support multiple testing frameworks and to allow the developer to specify the particular testing APIs to be used. This could be achieved using control codes (\ie special reserved keywords) to inform the model about the particular testing APIs used in the test case, both during training and inference.

\subsection{Deployment}
Deployment of large neural models to production represents a major engineering challenge. 
In this section, we discuss the possible deployment scenario in Visual Studio Code IDE backed by the Azure cloud compute.

We propose to design the \approach system as a two-layer service, consisting of the server-side inference module and the client-side unit test case provider module. With the model size exceeding 100 MB, the cloud-based deployment is the only viable option, which also offers control over the hardware setup and can guarantee resource availability. Introducing the client-side unit test case provider module would allow to minimize the inference time for the best user experience. The server-side module is deployed as a containerized web application to Azure Kubernetes Service~\cite{azurekubernetes} listening on a HTTPS endpoint. It processes completion requests and returns the model output, which is implemented in PyTorch.

%It is implemented in Python and executes model inference using PyTorch.
%\footnote{https://azure.microsoft.com/en-us/services/kubernetes-service/}

%Potential Discussion points:
%\begin{itemize}
%    \item Model Distillation
%    \item Potentially using Bart small/base
%    \item Cloud solution
%    \item incorporating this process in the CI pipeline
%\end{itemize}

\section{Threats to Validity}
Threats to \textit{construct validity} concern the relationship between theory and observation and are mainly related to the measurements we performed. In our context, the threat arises by training our models on potentially noisy data, specifically, low quality test cases or incorrect mapping between focal methods and tests. We attempt to mitigate this threat by relying on safe and accurate heuristics to mine test cases and focal methods, following best practices.

\textit{Internal validity} threats concern factors internal to our study
that could influence our results. The performance of our approach
depends on the hyperparameter configuration and pretraining process. We did not perform hyperparameter search since these large models require substantial training time, however, we reuse configurations suggested in the literature. We experiment with different pretraining stages and report the results of our experiments.

Threats to \textit{external validity} concern the generalization of our findings. In this paper the threat arises in RQ$_3$, given the small-scale evaluation, we cannot claim generalizability of the results. We clearly state that this represents a preliminary evaluation and more experiments should be conducted to assess the quality of our approach. We also acknowledge the fact that additional analyses should be performed to evaluate the fault detection capability of the generated test cases. We are actively working on addressing these limitations in our continuing work.

\section{Related Work}
Our work is related to several existing approaches in the area of automated software testing. In particular, there is a class of approaches that aims at generating tests cases, such as Evosuite \cite{fraser2011evosuite}, Randoop \cite{pacheco2007randoop}, and Agitar \cite{agitar}. The main differentiating factor between these techniques and our approach is the learning component. \approach is based on transformer model which aims at learning, from developer-written test cases, the best practices on how to write readable and accurate test cases. On the other hand, most of the existing techniques in the literature rely on handcrafted rules or heuristics to generate test cases, optimizing towards code coverage.

%Among these, Evosuite is one of the most popular tools for test generation in Java. It relies on mutation testing in order to generate appropriate assert statements. Specifically, it first introduces mutants within the method under test, then it attempts to generate asserts with the goal of killing the aforementioned mutants. During this process, Evosuite optimizes towards maximizing the number of killed mutants while generating as few asserts as possible. Randoop generates assert statements by relying on user-specified contracts. These statements are then refined using random testing and analyzing execution traces of the statement it creates.

%Deep learning applied to SE
Several existing works in the literature have proposed deep learning based approaches for software engineering tasks, such as code completion \cite{svyatkovskiy2019pythia}, automated patch generation \cite{tufano2019empirical, chen2019sequencer}, comment generation \cite{hu2018deep}, and many others \cite{watson2020learning}. While we share with these approaches the process of learning from examples, we also introduce significant novelty in this process. Specifically, we are among the first to train large, state-of-the-art sequence-to-sequence transformer models applied to software engineering tasks. Additionally, we pretrain these models on both English and source code showing the benefits of both types of pretraining on the generation of test cases.

%Recent advancements in language modelling
%Copied Alexey's section from Assert paper
Our work is also related to a broad set of literature on transfer learning~\cite{raffel2019exploring}, unsupervised language model pretraining~\cite{gpt,gpt2}, and denoising pretraining~\cite{bert, roberta, lewis2019bart}. In this paper, we extend these ideas to source code as a language, combining English and source code pretraining modes, fine-tuning on a downstream translation task from the automated software engineering domain. We compare this approach to the task-agnostic few-short learning approach introduced in GPT-3~\cite{brown2020language}. %Few-shot learning refers to a setting in which the model is given a few demonstrations of the task to complete at the inference time as conditioning. Unlike finetuning, no weight updates are performed in this case.
We find and discuss certain limitations of the few-shot learning approach as compared to finetuning using a translation task.

\section{Conclusion}
In this paper we presented \approach, an approach that aims at generating unit test cases by learning from real-world, developer-written test cases. Our approach relies on a sequence-to-sequence transformer model which was pretrained both on English and Java source code, then finetuned on the task of generating test cases given a method under test. We train the model using a supervised parallel corpus of 630k test cases and corresponding focal methods in Java, which we publicly release as \dataset~\cite{methods2test}.

Our evaluation shows that \approach is able to generate syntactically correct test cases that invoke a variety of testing APIs. We compiled and executed these test cases, comparing them with EvoSuite and GPT-3, finding that we achieve comparable or better test coverage. Finally, in a study with professional developers, we found that they prefer \approach's test cases in terms of readability, understandability, and testing effectiveness.

%We evaluate the ability of our model in generating test cases using natural language processing as well as code-specific criteria. First, we assess the quality of the translation compared to the target test case, then we analyze properties of the test case such as syntactic correctness and number and variety of testing APIs (\eg asserts). Finally, we execute the test cases, collect test coverage information, and compare them with test cases generated by EvoSuite and GPT-3.

\bibliographystyle{IEEEtran}
\balance
\bibliography{main}

%\begin{appendix}
%\section{Examples}
%\include{examples}
%\end{appendix}

\end{document}